\documentclass[twocolumn,pra,aps,floatfix,superscriptaddress]{revtex4-1}

\usepackage{amssymb,amsmath}
\usepackage{graphicx}
\usepackage{comment}
\usepackage{color}
\usepackage[colorlinks,urlcolor=blue,citecolor=blue,linkcolor=blue]{hyperref}
\usepackage{cleveref}
\usepackage{bigints}

\usepackage[normalem]{ulem}

\newcommand{\bra}[1]{\langle\left.{#1}\right|}
\newcommand{\ket}[1]{\left|{#1}\right.\rangle}
\renewcommand{\k}{\mathbf{k}}
\newcommand{\0}{\mathbf{0}}

\begin{document}

\title{Three-body correlations in a two-dimensional SU(3) Fermi gas}

\author{Thomas Kirk}
\affiliation{London Centre for Nanotechnology, 
17-19 Gordon Street, London, WC1H 0AH, United Kingdom}
\author{Meera M. Parish}
\affiliation{School of Physics and Astronomy, Monash University, Victoria 3800, Australia}

\date{\today}

\begin{abstract}
We consider a three-component Fermi gas that has SU(3) symmetry and is confined to two dimensions (2D). 
For realistic cold atomic gas experiments, we show that the phase diagram of the quasi-2D system can be characterized 
using two 2D scattering parameters: the scattering length and the effective range. 
Unlike the case in 3D, we argue that three-body bound states (trimers)  in the quasi-2D system can be stable against three-body losses. 
Using a low-density expansion coupled with a variational approach, we investigate the fate of such trimers in the many-body system as the attractive interactions are decreased (or, conversely, as the density of particles is increased). 
We find that remnants of trimers can persist in the form of strong three-body correlations in the weak-coupling (high-density) limit.
\end{abstract}

\pacs{}

\maketitle

\section{Introduction}
The three-component Fermi gas with short-range interactions is of fundamental interest owing to its resemblance to quark matter \cite{rapp2007,3-component_Background_Nishida,3-component_Background2_Nishida} and the possibility of exploring the interplay between pairing and three-body clustering. 
 Research into this system has gained further impetus from the recent experimental realization of three-component Fermi gases using ultracold $^6$Li atoms \cite{ottenstein2008collisional,FequencyAssociation_EfimovTriemrs_jochim,3-component_BindingEnergyMeasurement_Ueda,3-component_EfimovLoss_Jochim,3-component_RecombinationLoss_Hara,3-component_Efimov_RecombinationLoss2_Hara,Efimov_AtomDimer_Resonance_Jochim,Efimov_AtomDimer_Resonance_Ueda,Efimov_AtomDimer_Resonance_Ueda,3-component_BindingEnergyMeasurement_Ueda}.
 Thus far, experiments have successfully probed the \textit{few-body} properties  
of three distinguishable fermions --- most notably, Efimov trimers. These correspond to a series of 
three-body bound states (trimers) that exist when all the interactions are resonant, even in the absence of two-body bound states (dimers)~\cite{EfimovEffect_Efimov}.
The binding energy of the deepest bound Efimov trimer is set by the range of the interactions, while the energies of the weakly bound trimers 
obey a discrete scaling relation \cite{ScatteringConcepts_NaturalLength_Universality_Efimov_Brateen}. 
Efimov trimers were first observed experimentally 
for identical bosonic atoms, where they were detected as low-energy three-body loss resonances
\cite{EfimovObservation_Original_Kraemer}. 
Subsequently, the creation of three-component Fermi gases enabled the binding energies of Efimov trimers to be measured via radio-frequency association \cite{FequencyAssociation_EfimovTriemrs_jochim,3-component_BindingEnergyMeasurement_Ueda}.

Given the existence of universal few-body bound states in the three-component Fermi system, 
a natural question is how these will impact the many-body limit.
Much of the previous theoretical work has been restricted to two-body correlations described within mean-field theory \cite{Modawi1997,3-component_pairing_Paananen,3-component_FieldTheory_Symmetries_HighEnergy_He,3-component_BCSwavefunction_PhaseSeperation,3-component_finiteT_coexistence_Paananen,3-component_DomainWalls_Catelani,3-component_MediatedInteractions,3-component_BCSBECCrossover_Ozawa,Nummi2011,3-component_SpinOrbitCoupling_FFLO_Zhou}, where the focus was 
 on the crossover from BCS pairing to the Bose-Einstein condensation of dimers. 
However, when the possibility of trimers is explicitly included~\cite{3-component_Background_Nishida,3-component_RenomarlisationGroup_Floerchinger,3-component_3D_unbalanceda_Incao,niemann2012,3-component_Background2_Nishida,cui2015,azaria2009}, it is clear that three-body correlations cannot be neglected
outside of the weak-coupling regime.
Indeed, for sufficiently strong attraction, one expects a Fermi sea of trimer quasiparticles in the ground state~\cite{3-component_Background_Nishida,3-component_Background2_Nishida}. 
Nishida has further conjectured that the behavior of the three-component phase diagram is analogous to the quark-hadron continuity of nuclear matter, where fermionic quasiparticles change smoothly from atoms to trimers with increasing attraction
\cite{3-component_Background_Nishida}. This statement is supported by studies of a fermionic impurity in a paired-fermion superfluid~\cite{3-component_Background2_Nishida,cui2015}, but it remains to be seen whether such a crossover from atoms to trimers is stable against collapse~\cite{Blume2008}. 

Certainly, the three-body system is unstable towards decay into deeply bound states, thus making it an experimental challenge to create a stable trimer, let alone a many-body trimer phase.
The central issue is that atoms within the trimer tend to cluster together at short distances, particularly in the case of the deepest bound Efimov trimer, thus allowing two of the atoms to readily form a deeply bound dimer (which is absent in effective low-energy theories). 
Both the dimer and the remaining atom are then lost from the trapped system when the dimer binding energy is converted into kinetic energy.
Therefore, while it may be possible to realize a stable three-component  Fermi gas within a restricted parameter range~\cite{3-component_Efimov_RecombinationLoss2_Hara,Efimov_Lithium6_Theory_Hammer}, strong three-body losses are likely to preclude the existence of long-lived Efimov trimers  in current cold-atom experiments.

One proposed route to achieving stable trimers is to confine atoms to low dimensional geometries~\cite{Quasi2D_EfimovTrimers_Parish,yamashita2015}.
In particular, when identical bosons are restricted to move in two dimensions (2D), it has been shown that a short-range repulsion is present in the effective three-body potential~\cite{Quasi2D_EfimovTrimers_Parish}, thus suppressing losses \cite{3BodyRecombination_2D__Incao}. 
Indeed, in the strictly 2D limit, there is no Efimov effect and one has at most two universal trimers that are completely determined by the low-energy 2D scattering parameters~\cite{ThreeBosons_2D_Hammer,bruch1979binding}.
We expect the same situation to hold for the three-component Fermi system, since the trimers composed of identical bosons are identical to the spin-singlet trimers of fermions with SU(3) symmetry.

Furthermore, quasi-2D Fermi gases involving two spin components have already been successfully realized 
experimentally~\cite{2Dexperiments_pwavefermigas_Gunter,2Dgasobservation_fermions_Turlapov,2Dexperiment_fermions_Zwerger,From2Dto3D_Experimental_Vale,3-component_reference_Experimental_Quasi2D_Pairing_Experimental_Jochim,2Dexperiment_pairingpseudogap_Kohl,2Dexperiment_tightconfinement_Lithium6_Zwierlein,2Dexperiment_polarons_Thomas,2Dexperiment_Feshbachmolecules_Kohl,2Dexperiment_repulsivefermions,2Dexperiment_fermiliquids_kohl,2Dexperiment_fermiliquids_kohl,2Dexperiment_pressure_Turlapov}.
Motivated by the above,  
we will consider the phases of the \textit{two-dimensional} three-component Fermi gas in this paper. 
As far as we are aware, previous theoretical investigations of the 2D many-body system have been restricted to pairing phenomena~\cite{3-component_2D_FFLO_Chang}, and no three-body correlations were considered.  
For simplicity, we focus on SU(3) symmetry, where the masses, interactions and densities are the same for all species, but our approach can easily be extended to the imbalanced case. 

We will show how the quasi-2D system in cold-atom experiments can be described using effective 2D scattering parameters (the scattering length $a_{\rm 2D}$ and effective range $R_{\rm 2D}$) which are derived from the unidirectional confinement and the 3D scattering resonance. Moreover, we discuss how an effective 2D SU(3) model can approximately represent a realistic experimental system, and we provide evidence for the stability of three-body bound states in this model by estimating the three-body loss rates.
For the many-body system, we go beyond mean-field theory 
by precisely characterizing the low-density (few-body) limit and by employing a variational wave function to investigate three-body correlations in the high-density regime. On the basis of this, we argue that stable three-body bound states can evolve into strong three-body correlations with increasing particle density.

The manuscript is organized as follows. Section \ref{model} outlines our effective 2D model for a three-component Fermi gas confined to a quasi-2D geometry, explaining its connection to the two-body physics. In Sec.~\ref{three-body}, we determine the three-body bound states that exist within our 2D model, and we estimate their size and stability as a function of the scattering parameters $a_{\rm 2D}$ and $R_{\rm 2D}$. 
In Sec. \ref{many-body} we analyse the many-body problem by considering  different limits of the phase diagram.
Firstly, we address the low-density regime using a perturbative expansion based on the few-body states of 
Sec.~\ref{three-body}. Secondly, we employ mean-field theory to obtain the leading order behavior in the limit of large effective range. Finally, we investigate the high-density limit using a variational ansatz. We conclude in Sec.~\ref{sec:conc}.

\section{The Quasi-2D system}\label{model}
In this section, we will motivate and describe an effective 2D model of a confined ultracold atomic gas experiment. Before stating our effective 2D Hamiltonian, let us introduce our model's key parameters by relating them to  familiar experimental quantities. We make this connection using two-body scattering theory. Throughout the manuscript, we set $\hbar=1$.

In a real experiment, one constructs a quasi-2D system by strongly confining a 3D atomic gas along one direction.
Therefore, let us first consider the 3D scattering properties of a three-component Fermi gas.
In the  $^6$Li Fermi gas 
 experiments \cite{ottenstein2008collisional,FequencyAssociation_EfimovTriemrs_jochim,3-component_BindingEnergyMeasurement_Ueda,3-component_EfimovLoss_Jochim,3-component_RecombinationLoss_Hara,3-component_Efimov_RecombinationLoss2_Hara,Efimov_AtomDimer_Resonance_Jochim,Efimov_AtomDimer_Resonance_Ueda,Efimov_AtomDimer_Resonance_Ueda,3-component_BindingEnergyMeasurement_Ueda}, 
 the scattering between two distinguishable atoms (i.e., two different hyperfine states of lithium)
 is low in energy and is well described by the  $s$-wave two-body scattering amplitude: 
\begin{equation}
f_{\rm 3D}(E) = \frac{1}{-a_{\rm 3D}^{-1}+\sqrt{-mE -i0} - m R_{\rm 3D} E} \ ,
\end{equation}
where $E$ is the relative energy of the two atoms, each with mass $m$, 
while  $a_{\rm 3D}$ and $R_{\rm 3D}$ are the scattering length and range parameter, respectively. The real part of the denominator is obtained from a low-energy expansion of the scattering phase shift, 
which is appropriate for systems in the low-energy regime. Experimentally, the scattering length $a_{\rm 3D}$ can be varied using a magnetically tunable Feshbach resonance~\cite{FeshbachResonanca_Review_Grimm}, and the range parameter $R_{\rm 3D}$ is determined by the width of the resonance, such that we have $R_{\rm 3D} >0$.
The location of the resonance, $a_{\rm 3D}^{-1}=0$,  and the value of $R_{\rm 3D}$ are fixed for a given pair of hyperfine states. Here, we will assume that they are the same for all pairs of hyperfine states in the three-component Fermi gas. 
This is not such a drastic assumption since there are overlapping broad Feshbach resonances in the $^6$Li system \cite{3-component_reference_Experimental_FeshbachRsonancesLithium_ScatteringLength_Bartenstein}, and
we still expect our results to be qualitatively correct when the scattering lengths and range parameters are slightly different.

Now consider applying a harmonic confining potential along the $z$ direction with angular frequency $\omega_z$, i.e., $V(z)  = \frac{1}{2} m\omega_z^2 z^2$.  
By expressing the two-body problem in the basis of harmonic oscillator levels in the $z$ direction, one can derive a quasi-2D scattering amplitude that is a function of the relative wavevector $\k$ between two atoms in the 2D plane \cite{PhysRevA.64.012706,Q2DBackground_Review_Meera}.
In the limit of strong confinement $\omega_z \gg k^2/m$, with $k \equiv |\k|$, 
we can expand the quasi-2D scattering amplitude to obtain \cite{2Dexperiment_Feshbachmolecules_Kohl,PhysRevA.64.012706,Q2DBackground_Review_Meera}:
\begin{equation} \notag
f_{\rm q2D}(k) \simeq \frac{4\pi}{i\pi+\frac{\sqrt{2\pi}l_z}{a_{\rm 3D}}-\ln(\frac{\pi}{B}k^2l_z^2)+\frac{\sqrt{2\pi}R_{\rm 3D}}{l_z}(k^2l_z^2+\frac{1}{2})},
\end{equation}
where $B=0.905$ and the confinement length $l_z=\sqrt{\frac{1}{m\omega_z}}$. 
Equivalently, one can view this as an expansion in small $k l_z$, where we have kept terms in the denominator up to order $l_z$. For this expansion to be valid, we also require $R_{\rm 3D} \gtrsim l_z$ and $l_z \gg v_{\rm vdW}$, where $v_{\rm vdW}$ is the van der Waals range of the interactions between atoms.
For a sufficiently broad resonance where $R_{\rm 3D} \ll l_z$, we can set $R_{\rm 3D}$ to zero in $f_{\rm q2D}(k) $.

In the regime of strong confinement, a purely 2D model  is sufficient for describing the behaviour of the Fermi system since the two-body scattering of distinguishable fermions maps onto the 2D scattering amplitude \cite{2D_scatteringtheory_Adhikari}:
\begin{equation} \label{eq:2dscat}
f_{\rm 2D}(k) = \frac{4\pi}{i\pi-\ln(k^2a_{\rm 2D}^2)+R_{\rm 2D}^2k^2},
\end{equation}
where $a_{\rm 2D}$ is the 2D scattering length and $R_{\rm 2D}$ is the 2D effective range. 
Thus, using  $f_{\rm q2D}(\k)$,  one can relate the familiar 3D quantities $l_z$, $a_{\rm 3D}$ and $R_{\rm 3D}$ to the purely 2D scattering parameters $a_{\rm 2D}$ and $R_{\rm 2D}$:
\begin{align}\label{eq:from2dto3d}
\begin{split}
  a_{\rm 2D}   &= l_z\sqrt{\frac{\pi}{\tilde{B}}}e^{-\sqrt{\frac{\pi}{2}}\frac{l_z}{a_{\rm 3D}}} \\
  R_{\rm 2D}^2 &= \sqrt{2\pi}R_{\rm 3D}l_z,
\end{split}
\end{align}
where $\tilde{B}=B\exp(\sqrt{\frac{\pi}{2}}R_{\rm 3D}/l_z)$. Note that the 2D scattering length is modified from the usual expression (see, e.g., Ref.~\cite{Q2DBackground_Review_Meera}) due to the fact that the 3D effective range couples to the zero-point energy in the quasi-2D geometry. 

With this connection in mind, we proceed to write down an effective 2D Hamiltonian for the three-component Fermi gas which captures the required two-body scattering behavior in the strongly confined limit.

\subsection{Effective 2D model}
The effective 2D model that generates the scattering amplitude \eqref{eq:2dscat}  
contains three species of fermions  interacting via closed-channel stuctureless bosons. 
Labelling the different fermion flavors by $i=1,2,3$, we have the following Hamiltonian (with system area set to 1): 
\begin{multline}\label{hamiltonian}
\hat{H} -\mu \hat{N} -2\mu \hat{N}_{b} = \\
\sum_{\mathbf{k},i}(\epsilon_{\mathbf{k}}-\mu)c^{\dagger}_{\mathbf{k},i}c_{\mathbf{k},i}+\sum_{\mathbf{k},i}\left(\frac{1}{2}\epsilon_{\mathbf{k}}+\nu-2\mu \right)b^{\dagger}_{\mathbf{k},i}b_{\mathbf{k},i}\\ 
+\frac{g}{2}\sum_{\mathbf{k},\mathbf{q}}\sum_{i,j,l}\left(\epsilon_{ijl}\, b_{\mathbf{q},i}c_{\mathbf{k},j}^{\dagger}c_{\mathbf{q}-\mathbf{k},l}^{\dagger}+h.c.\right).
\end{multline}
Here, $c^{\dagger}_{\mathbf{k},i}$ creates a type-$i$ fermion with momentum $\mathbf{k}$, while $b^{\dagger}_{\mathbf{k},i}$ creates a closed-channel boson with momentum $\mathbf{k}$ and flavor $i$ 
(which is distinct from a fermion of type $i$). 
Each pair of fermions is coupled to a closed-channel boson via the interaction coefficient $g$. 
Note that there are three different types of bosons corresponding to the three distinct pairs of fermions, as encapsulated by the Levi-Civita tensor $\epsilon_{ijl}$. 
The kinetic energy $\epsilon_\k =k^2/2m$, while $\mu$ is the chemical potential of the system, and $\nu$ is the \textit{bare} detuning between the open and closed channels.
Since the open and closed channels are coupled by the Hamiltonian $\hat{H}$, we can assume that the total number of fermions $N$ is in chemical equilibrium with the total number of closed-channel bosons $N_b$.

Following Ref.~\cite{two-component_fermions_Gurarie},
the scattering physics of Eq.~\eqref{eq:2dscat} can be connected to the two-channel
model by relating the bare parameters $g$ and $\nu$ to the measurable quantities of the scattering amplitude, $a_{\rm 2D}$ and $R_{\rm 2D}$:
\begin{align}\label{renorm}
\begin{split}
  a_{\rm 2D}   = \frac{1}{\Lambda}e^{\frac{2\pi\nu}{mg^2}}, & \hspace{15pt}
  R_{\rm 2D}^2 = \frac{4\pi}{m^2g^2},
\end{split}
\end{align}
where $\Lambda$ is a ultraviolet cutoff such that all momenta $|\k| < \Lambda$. In what follows, we always take limit $\Lambda \to \infty$, while keeping the scattering parameters finite. Note that this means that we also require $\nu \to \infty$.

\subsection{Two-body bound state in 2D} 
The 2D scattering parameters can, in turn, be related to a two-body bound state, which always exists in 2D \cite{landlandau}. Such a bound state corresponds to a pole of the 2D scattering amplitude, i.e., for a dimer with energy $\varepsilon = k^2/m < 0$, the (imaginary) value of $k$ satisfies the condition $f_{\rm 2D}(\mathbf{k})^{-1}=0$.
Equivalently, one can obtain the dimer energy by considering the action of the Hamiltonian on the dimer wave function, as in the following.

Within our effective 2D model, a general two-body state is a superposition of a closed-channel boson and a pair of two distinguishable fermions. 
Due to the translational invariance of the system, the center-of-mass motion decouples from the relative pair motion, and we can thus set the center-of-mass momentum to zero without loss of generality.
Taking the example of a 2-3 pair of fermions, the two-body wave function is:
\begin{equation}\label{2body}
|\Psi_2\rangle = \alpha b^{\dagger}_{\mathbf{0},1}|0\rangle +\sum_{\mathbf{k}}\beta_{\mathbf{k}}c^{\dagger}_{\mathbf{k},2}c^{\dagger}_{-\mathbf{k},3}|0\rangle
\end{equation}
where $|0\rangle$ is the vacuum state, and the amplitudes $\alpha$, $\beta_\k$ satisfy: $|\alpha|^2 + \sum_\k |\beta_\k|^2 =1$.  Due to the SU(3) symmetry,
our results for the dimer do not depend on which pair of distinguishable fermions we choose. In particular, we can construct a dimer with equal components of the fermions species $i$ by transforming to the new basis:
\begin{align} \label{eq:transform}
\begin{pmatrix}
c_{\k,1'} \\
c_{\k,2'} \\
c_{\k, 3'}
\end{pmatrix}
= 
\begin{pmatrix}
1 & 1 & 1 \\
1 & e^{i2\pi/3} & e^{i4\pi/3} \\
1 & e^{-i2\pi/3} & e^{-i4\pi/3} 
\end{pmatrix}
\begin{pmatrix}
c_{\k,1} \\
c_{\k,2} \\
c_{\k, 3}
\end{pmatrix} 
\end{align}
and then considering, e.g., a $2'$-$3'$ pair.

Using the Hamiltonian from Eq.~\eqref{hamiltonian} and solving the eigenvalue equation $\hat{H} |\Psi_2\rangle = E_2 |\Psi_2\rangle$,
we obtain an expression for the vacuum dimer energy $E_2$:
\begin{equation}
\frac{\nu -E_2}{g^2}=\sum^{\Lambda}_{\mathbf{k}}\frac{1}{2\epsilon_{\mathbf{k}} -E_2}.
\end{equation}
Inserting the relations \eqref{renorm}, one obtains the cutoff independent
solution~\cite{ThreeBosons_2D_Hammer}:
\begin{equation}	\label{2bod}
E_2=- \frac{1}{mR_{\rm 2D}^2} W\bigg(\frac{R_{\rm 2D}^2}{a_{\rm 2D}^2}\bigg),
\end{equation}
where $W(x)$ is the Lambert-W function. Note that 
there is always one two-body bound state
for a given $a_{\rm 2D}$ and $R_{\rm 2D}$.
When the 2D effective range is zero, we recover the well-known result $E_2 = -1 /m a_{\rm 2D}^2$ \cite{PhysRevLett.62.981}, where $a_{\rm 2D}$ corresponds to the size of the dimer in this case. 
The behavior of $E_2$ for general $R_{\rm 2D}/a_{\rm 2D}$ is displayed in Fig.~\ref{2and3fig}, where we have plotted the binding energy per atom $|E_2|/2$. In the limit $R_{\rm 2D} \ll a_{\rm 2D}$, we obtain $m|E_2| \simeq 1/(a_{\rm 2D}^2 + R_{\rm 2D}^2)$, while in the opposite limit $R_{\rm 2D}/a_{\rm 2D} \to \infty$, we have $m R^2_{\rm 2D} |E_2| \to 2\ln(R_{\rm 2D}/a_{\rm 2D})-2\ln\big(\ln(R_{\rm 2D}/a_{\rm 2D})\big)$. To remain in the 2D regime of the quasi-2D geometry, we require $|E_2| \ll \omega_z$, such that the dimer is flattened out within the plane and the excited levels of the confining potential can be neglected~\cite{Q2DBackground_Review_Meera}.

\section{Three-body bound states}\label{three-body}
We now turn to the three-body problem within our 2D SU(3) model and determine the properties of the bound three-body \textit{trimer} states. We will see that the trimers of our model are identical to
those which occur in a system of three indistinguishable bosons \cite{ThreeBosons_2D_Hammer,bruch1979binding}. 
Note, however, that the situation is different for systems of more than three particles, since the three-component Fermi system does not have any $s$-wave $N$-body bound states for $N>3$, in contrast to the single-component Bose system. The results of this section will show that 
strong three-body correlations cannot be ignored when determining the many-body state of the SU(3) Fermi gas.

\begin{figure}\centering
  \includegraphics[width=0.9\columnwidth]{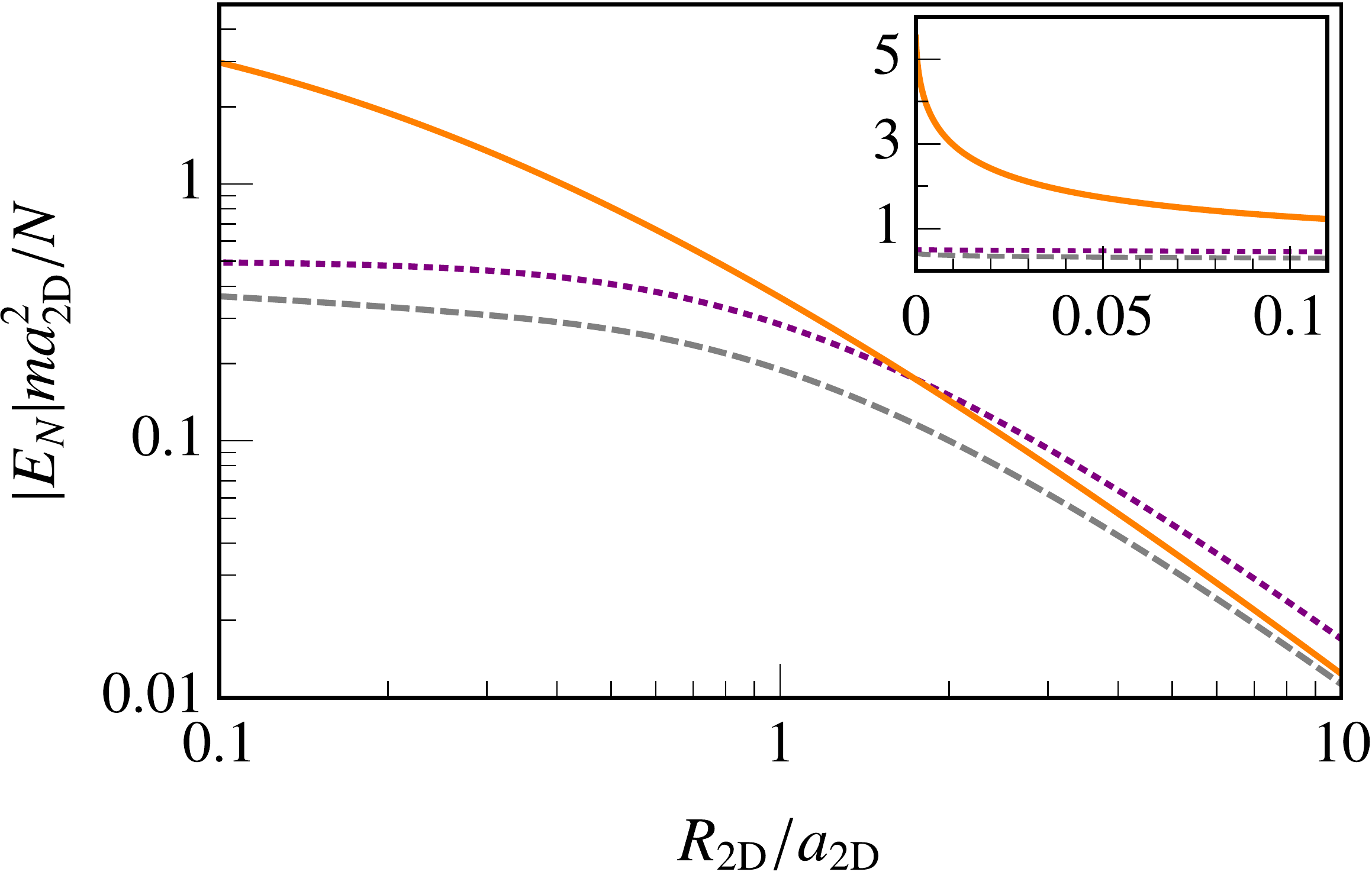}%
 \caption{
 Binding energy per atom of different few-body states. This corresponds to $-\mu$ of the many-body system in the limit of vanishing density (see text).
 The solid (orange) line is the ground-state trimer of the three-body problem, the dashed (grey) line is the excited trimer state of the three-body problem, 
 and the dotted (purple) line is the dimer state of the two-body problem. At zero effective range, the ground-state and excited-state trimer energies are $E_3 m a_{\rm 2D}^2/3 \approx -5.5$ and $-0.42$, respectively. The inset shows the same curves with linear scales on the axes in order to expose the behavior as $R_{\rm 2D}/a_{\rm 2D}\rightarrow 0$.} \label{2and3fig}
\end{figure}

Similarly to the two-body problem, we can determine the bound states of three distinguishable fermions by considering a wave function containing all possible combinations of fermions and closed-channel bosons at zero center-of-mass momentum:
\begin{multline}\label{3wave}
|\Psi_3\rangle = \sum_{\mathbf{k},i}\alpha_{\mathbf{k},i}b^{\dagger}_{\mathbf{k},i} c^{\dagger}_{-\mathbf{k},i}|0\rangle \\
+\sum_{\mathbf{k}_1,\mathbf{k}_2,\mathbf{k}_3}\beta_{\mathbf{k}_1 \mathbf{k}_2 \mathbf{k}_3} \, \delta (\mathbf{k}_1+\mathbf{k}_2+\mathbf{k}_3) \, c_{\mathbf{k}_1,1}^{\dagger}c_{\mathbf{k}_2,2}^{\dagger}c_{\mathbf{k}_3,3}^{\dagger}|0\rangle .
\end{multline}
By acting our 2D Hamiltonian on this wave function, we obtain the Schr\"odinger equation for the three-body energy: $\hat{H}|\Psi_3\rangle=E_3|\Psi_3\rangle$. This leads to a set of four coupled equations: 
\begin{align} \notag
  E_3 \alpha_{\mathbf{k},i}   &=  \left(\frac {3}{2}\epsilon_{\mathbf{k}}+\nu\right)\alpha_{\mathbf{k},i}
+ \eta_{\mathbf{k}}^{(i)} 
\\ \notag
E_3\beta_{\mathbf{k}_1\mathbf{k}_2\mathbf{k}_3}
&= \left(\epsilon_{\mathbf{k}_1}+\epsilon_{\mathbf{k}_2}+\epsilon_{\mathbf{k}_3}\right)\beta_{\mathbf{k}_1\mathbf{k}_2\mathbf{k}_3} 
+g\sum_{i=1}^3 \alpha_{\mathbf{k}_i,i},
\end{align}
where we have the restriction $\k_1 +\k_2 +\k_3 = 0$ in the last equation, 
and we have defined the functions
\begin{align} \label{eq:eta}
\eta^{(i)}_{\k_i} = g \sum_{\{\k_j\}_{j\neq i}}
\beta_{\mathbf{k}_1\mathbf{k}_2\mathbf{k}_3} \delta(\mathbf{k}_1+\mathbf{k}_2+\mathbf{k}_3) \, .
\end{align}
Eliminating the amplitudes $\beta_{\mathbf{k}_1\mathbf{k}_2\mathbf{k}_3}$ reduces the problem to three coupled equations:
\begin{align}\notag
&\left[\frac{1}{g^2}\left(E_3-\nu-  \frac{3}{2}\epsilon_{\mathbf{k}}\right) -\sum_{\mathbf{k}'}\frac{1}{E_3-\epsilon_{\mathbf{k}}-\epsilon_{\mathbf{k}'}-\epsilon_{\mathbf{k}-\mathbf{k}'}} \right]\alpha_{\mathbf{k},i}
\\
&\hspace{22mm} =\sum_{\mathbf{k}'}\sum_{j\neq i}\frac{\alpha_{\mathbf{k}',j}}{E_3-\epsilon_{\mathbf{k}}-\epsilon_{\mathbf{k}'}-\epsilon_{\mathbf{k}+\mathbf{k}'}}.
\end{align}
We can further reduce these down to a single integral equation by rewriting everything in terms of the 
function $C_{\mathbf{k}}=\sum_{i=1}^3\alpha_{\mathbf{k},i}$ and then using \eqref{renorm} to regularise it: 
\begin{align} \notag
&\left[mR_{2D}^2\left(E_3-\frac{3}{2}\epsilon_{\mathbf{k}}\right)-\ln \left(ma_{2D}^2\left(-E_3+\frac{3}{2}\epsilon_{\mathbf{k}}\right)\right)\right]C_{\mathbf{k}}
\\ \label{3bod}
&\hspace{22mm} =\frac{8\pi}{m}\sum_{\mathbf{k}'}\frac{C_{\mathbf{k}'}}{E_3-\epsilon_{\mathbf{k}}-\epsilon_{\mathbf{k}'}-\epsilon_{\mathbf{k}+\mathbf{k}'}}.
\end{align}
Importantly, this equation is 
equivalent to that found with diagrammatic methods in the problem of three identical bosons~\cite{ThreeBosons_2D_Hammer,bruch1979binding}.
Furthermore, while one can construct two more decoupled equations involving other linear combinations of $\alpha_{\mathbf{k},i}$, Eq.~\eqref{3bod} is the only equation that yields three-body bound states.
Thus, we can simply set $\alpha_{\mathbf{k},i} = C_{\mathbf{k}}/3$ when considering trimer states.

Solving Eq.~\eqref{3bod} allows us to determine the energies $E_3$ of the bound trimers.
 At zero effective range, there are two trimer states, ground and excited, which have the universal energies $-16.5/ma_{\rm 2D}^2$ and $-1.27/ma_{\rm 2D}^2$, respectively~\cite{bruch1979binding}. 
 In Fig.~\ref{2and3fig}, we depict the binding energy per atom, $|E_3|/3$, of these two trimers as a function of the effective range. 
As we discuss in Sec.~\ref{sec:low}, the energy per atom (as determined from few-body calculations) 
can correspond to the chemical potential $\mu$ of the SU(3) many-body system in the limit of vanishing density.
For instance, $\mu = E_3/3$ is the chemical potential of an extremely dilute gas of bound trimers, while $\mu = E_2/2$ is that of a gas of dimers.
We see in Fig.~\ref{2and3fig} that the binding energy per fermion (corresponding to $-\mu$) for the excited trimer is always lower than that of the dimer and is thus never stable in the many-body system. On the other hand, the ground-state trimer is stable up until the critical effective range $R_{\rm 2D}/a_{\rm 2D} \simeq 1.7 $.  

Unlike the case in 3D, one can show that there is \emph{always} a 2D bound trimer state in the \emph{three}-body system. To see this, we consider the unbinding transition of a trimer into a dimer and an atom, where $E_3 \to E_2$ and $C_\k \to \delta_{\k,\0} \, C_\0$. Rewriting Eq.~\eqref{3bod} so that we absorb the bracketed term on the \textit{l.h.s.} into $C_\k$ to get $C^\prime_\k$, we then obtain for the $\k = \0$ component
\begin{align}
 \frac{3m (1 - m E_2 R^2_{\rm 2D})} {16\pi E_2} C^\prime_\0 \simeq \sum_{\mathbf{k}'}
    \frac{C^\prime_{\mathbf{k}'}}{\epsilon_{\mathbf{k}'}(E_2-2\epsilon_{\mathbf{k}'})},
\end{align}
where we have taken $E_3 = E_2$ and we have considered small momenta, $|\mathbf{k}'| \ll \sqrt{m|E_2|}$. 
Converting the sum to a 2D integral, we find that the \textit{r.h.s.} diverges logarithmically for $\k' \to \0$. Thus, the condition for unbinding can only be satisfied in the limit $R_{\rm 2D} \to \infty$, which demonstrates that the trimer is always bound in the three-body system. 

\subsection{Lifetime of the spin-singlet trimer in 2D}

\begin{figure}
\centering
 \includegraphics[width=0.9\columnwidth]{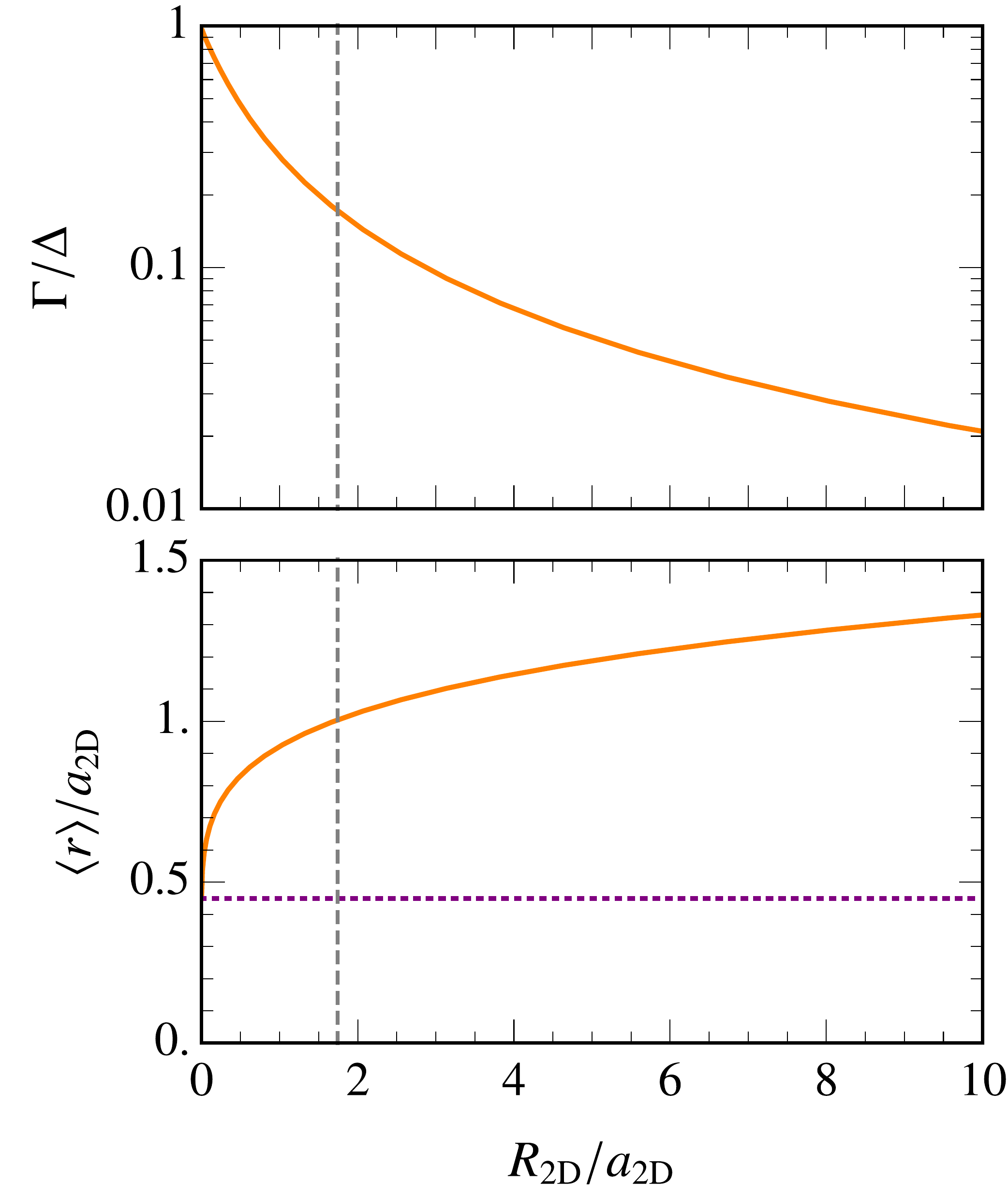} 
\caption{
The solid (orange) lines show the rescaled decay rate $\Gamma/\Delta$ (top) and size $\langle r\rangle$ (bottom) of the ground-state trimer. The dotted (purple) lines show the corresponding values at zero effective range whilst the dashed (grey) line shows the critical effective range beyond which a dilute gas of trimers is unstable towards forming dimers. 
}
\label{sizeanddecay}
\end{figure}

The existence of three-body losses in experiment makes it a challenge to realise long-lived trimers.
For the three-component Fermi gas in 3D, the Efimov trimers are extremely unstable because their wave functions have a significant overlap with a deeply bound dimer state \cite{ScatteringConcepts_NaturalLength_Universality_Efimov_Brateen}, thus allowing a pair of atoms to readily transition into the deeper dimer state. 
 The large dimer binding energy is then converted into kinetic energy of the remaining atom and dimer, 
 which causes the particles to escape from the trap.
Indeed, most of the experimental evidence for the Efimov trimer relies on measurements of the loss rate of atoms from the trapped gas~\cite{EfimovObservation_Original_Kraemer,3-component_RecombinationLoss_Hara,3-component_Efimov_RecombinationLoss2_Hara,Efimov_AtomDimer_Resonance_Ueda,3-component_EfimovLoss_Jochim}.

To suppress three-body losses, one must therefore prevent three atoms from clustering close together within the trimer state. This can be achieved by having a repulsive barrier at short distances in the effective three-body potential~\cite{Quasi2D_EfimovTrimers_Parish,LongLiveTrimers_2-component_Jesper,0034-4885-80-5-056001}.
 For instance, in a system of two heavy fermions and one light particle, there is a short-range centrifugal barrier in the three-body potential due to the Pauli exclusion between the identical fermions. In 3D, this leads to long-lived universal trimers, provided that the heavy-light mass ratio is less than $\sim 13.6$ \cite{KartavtsevMalykh_Trimers_Kartavtsev,Efimov_KartavtsevMalykh_CrossoverTrimers_Shimpei}.
 
In the absence of Pauli exclusion, short-range repulsion in the three-body system can also be engineered by confining the atoms to a quasi-2D geometry~\cite{Quasi2D_EfimovTrimers_Parish,3BodyRecombination_2D__Incao,ThreeBosons_2D_Hammer}. 
As shown for three identical bosons, a tight confinement along the $z$ direction produces an attractive well at long distances in the three-body potential, creating extended quasi-2D trimers with a small weight at short distances \cite{Quasi2D_EfimovTrimers_Parish}. Such 2D trimers are thus expected to have a reduced rate of three-body recombination. 
 
As seen from Eq.~\eqref{3bod}, the trimers in the SU(3) three-component Fermi system are the same as those for three identical bosons. Therefore, we can, in the same manner, 
exploit confinement to stabilize the three-component Fermi system. 
Furthermore, our model contains an additional tuning parameter -- the range $R_{\rm 3D}$ -- which is absent in previous studies of the quasi-2D $s$-wave trimer \cite{Quasi2D_EfimovTrimers_Parish} (although it has been considered for heteronuclear $p$-wave trimers in a quasi-2D geometry~\cite{LongLiveTrimers_2-component_Jesper}).
Confinement can thus affect the few-body physics via both $a_{\rm 2D}$ and $R_{\rm 2D}^2 \sim l_z R_{\rm 3D}$ in the quasi-2D scattering amplitude. We now proceed to analyse how the confinement impacts the lifetime of the trimers 
for a general $R_{\rm 3D}$.

We wish to describe the short-range decay process and subsequent escape of atoms from the experimental trap. To capture this non-unitary loss of atoms, we consider  
a non-Hermitian
perturbation to our Hamiltonian~\cite{AtomDimer_DimerDimerScattering_Jesper} 
(see App.~\ref{decayratederivation}). The specific form of the perturbation must model situations in which the three fermionic atoms are all close together within the trimer. Since the closed-channel boson effectively contains two atoms at zero separation, we chose the following perturbation:
\begin{equation}\label{perturbation}
\hat{H}_{3b}=-i \frac{\Delta}{2} \sum_{\mathbf{k},\mathbf{k}',\mathbf{q},i} b^{\dagger}_{\mathbf{q}-\mathbf{k}',i}c^{\dagger}_{\mathbf{k}',i}c_{\mathbf{k},i}b_{\mathbf{q}-\mathbf{k},i} \, ,
\end{equation}
where $\Delta$ is a real coupling constant. This describes the scattering of a 
closed-channel dimer and an atom at short distances.
Such a loss process has been used in Ref.~\cite{jag2016} to successfully model the losses  in $^6$Li-$^{40}$K mixtures.
For the SU(3) system, the loss rate $\Gamma$ due to this process is given by:
\begin{align} 
\Gamma = 2i 
\langle \Psi_3| \hat{H}_{3b} |\Psi_3\rangle 
= \frac{\Delta}{3}\Big|\sum_{\mathbf{k}}C_{\mathbf{k}}\Big|^2 \, ,
\end{align}
where $C_{\mathbf{k}}$ is found from the integral equation \eqref{3bod}. The decay rate is shown in 
Fig.~\ref{sizeanddecay} as a function of the 2D effective range, $R_{\rm 2D}/a_{\rm 2D}$. Using 
Eq.~\eqref{eq:from2dto3d}, we can express this dimensionless parameter in terms of experimental quantities as follows:
\begin{equation}\label{growth}
R_{\rm 2D}^2/a_{\rm 2D}^2 = \sqrt{\frac{2}{\pi}} B \frac{R_{\rm 3D}}{l_z}  e^{\sqrt{\frac{\pi}{2}}\frac{R_{\rm 3D}}{l_z}+\sqrt{2\pi}\frac{l_z}{a_{\rm 3D}}} \, ,
\end{equation}
where $a_{\rm 3D} < 0$ for a trimer in the 2D limit. Therefore, for increasing confinement (decreasing $l_z$), the ratio $R_{\rm 2D}/a_{\rm 2D}$ becomes larger and the decay rate decreases. This reveals a new mechanism by which the trimer decay rate in an ultracold atomic gas experiment may be suppressed by a confining potential.

This increased lifetime of the trimers has a simple explanation: The spatial size of the trimer increases with increasing $R_{\rm 2D}/a_{\rm 2D}$, such that the short-range decay events are less likely to occur. To see this, we estimate the 
spatial size of the trimer 
using the real-space wave function $\psi(\mathbf{r})=\mathcal{N}^{-1/2}\sum_{\mathbf{k}} e^{i\mathbf{k}.\mathbf{r}}C_{\mathbf{k}}$, where $\mathcal{N}$ is the normalization factor such that  $\int  \left|\psi(\mathbf{r})\right|^2d\mathbf{r} = 1$. This wave function represents one particular configuration of the trimer state, 
where the separation between one pair of the particles is set to zero (see App.~\ref{sizeappendix}). The variable $\mathbf{r}$ corresponds to the real-space separation between the point-like pair and the third particle. The expectation value of the distance $r\equiv|\mathbf{r}|$ is readily calculated as follows:
\begin{align} \label{eq:expect}
\langle r\rangle&=\int r \big|\psi(\mathbf{r})\big|^2 d\mathbf{r} \, .
\end{align}
This quantity is plotted in the lower part of Fig.~\ref{sizeanddecay}. We see how the size of the trimer grows with increasing $R_{\rm 2D}/a_{\rm 2D}$ and therefore with increasingly strong confinement, thus suppressing short-range three-body losses and extending the lifetime of the trimer.

\section{Many-body system}\label{many-body}

\begin{figure}
 \centering
 \includegraphics[width=0.9\columnwidth]{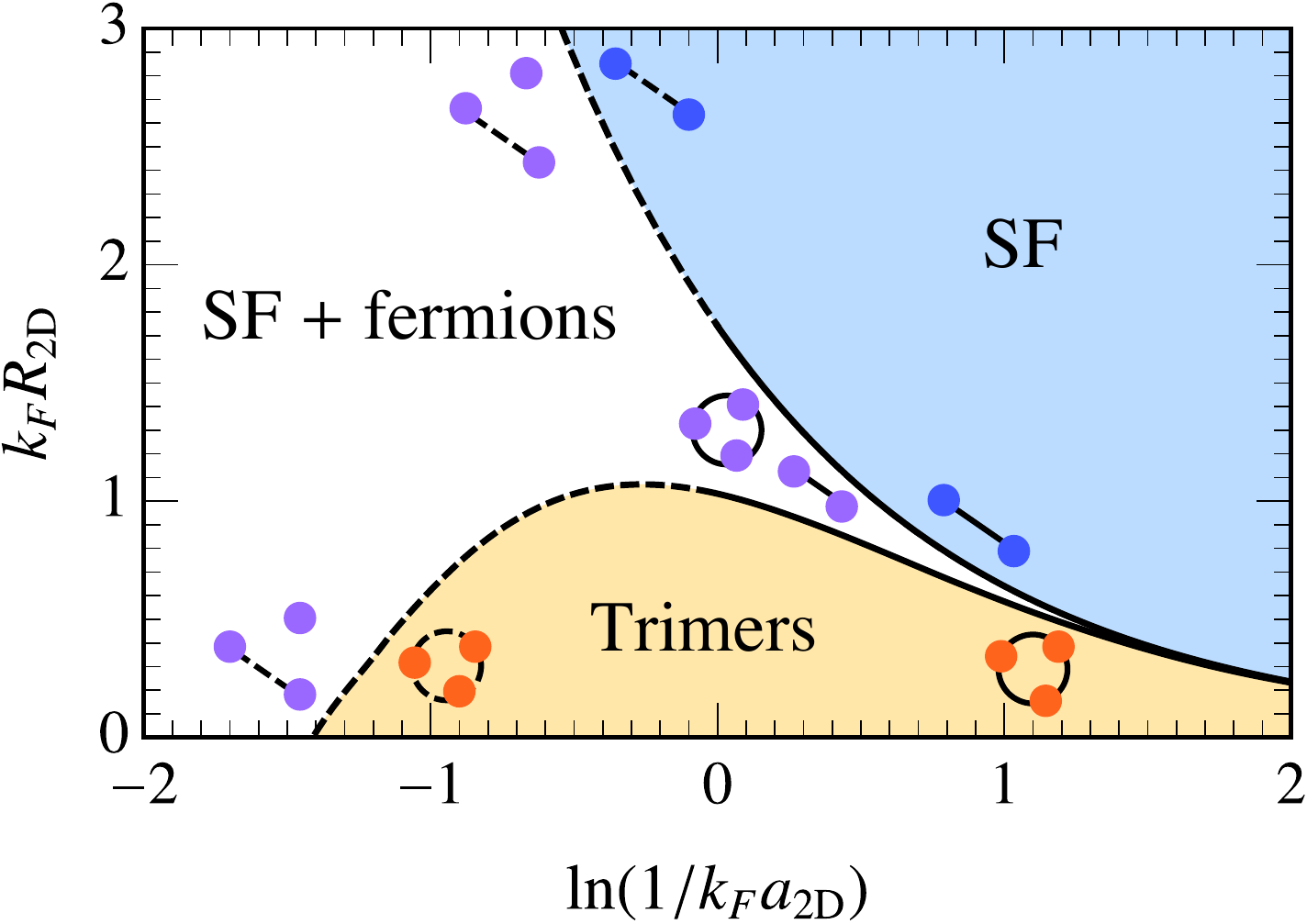}
 \caption{Schematic phase diagram of the 2D SU(3) Fermi gas as a function of the effective range and scattering length.
 The blue SF region designates the fully paired superfluid phase, the orange shaded region corresponds to trimers only, and the remaining area has a mixture of a superfluid and fermions (trimers or unbound atoms).  
 In the low-density limit, where $\ln(1/k_F a_{\rm 2D}) \gg 1$ and $k_F R_{\rm 2D} \ll 1$, the system corresponds to a weakly interacting gas of dimers and/or trimers, as illustrated by the particle clusters linked by solid lines. Here, when $R_{\rm 2D} \to 0$, the ground state is a trimer Fermi gas, while increasing $k_F R_{\rm 2D}$ yields a transition to a trimer-dimer mixture and then finally a dimer-only superfluid. In the limit $k_F R_{\rm 2D} \gg 1$, there is a transition from the fully paired superfluid to weak BCS pairing plus unbound atoms as we decrease $\ln(1/k_F a_{\rm 2D})$. 
 In the opposite regime $k_F R_{\rm 2D} < 1$, there is a region defined by strong three-body correlations (illustrated by orange particles linked with dashed lines), where the size of the three-body clusters can be comparable to or larger than the interparticle spacing when $\ln(1/k_F a_{\rm 2D}) \lesssim -1$.
Eventually, for sufficiently weak attraction, these three-body clusters are replaced by Cooper pairs coexisting with an atomic Fermi sea, which smoothly connects with the high-density phase in the $k_F R_{\rm 2D} \gg 1$ limit.}
 \label{PhaseDiagram}
\end{figure}

We now turn to the phase diagram of the many-body system at zero temperature.
Here, we consider a three-component SU(3) Fermi gas in 2D with area density $n$ for each species of fermion (corresponding to total density $3n$).
We parametrize the many-body problem with the quantities $k_F a_{\rm 2D}$ and $k_F R_{\rm 2D}$, where we have defined the Fermi momentum $k_F = \sqrt{4 \pi n}$.
We tackle the phase diagram of the many-body problem 
by calculating three limits: Firstly, we characterize the low-density regime $k_F \to 0$,
where we can use our few-body results as a basis for a  low-density perturbative analysis. Secondly, we investigate the 
regime $k_F R_{\rm 2D}\gg1$, where BCS mean-field theory should capture the leading order behavior in $1/k_F R_{\rm 2D}$ \cite{two-component_fermions_Gurarie}. Finally, we analyse the high-density, weak-coupling regime $k_F a_{\rm 2D} \gg 1$ by using a variational ansatz that incorporates strong three-body correlations in the Fermi gas. The approximations we make in each limit are 
designed to extract the essential physics, without recourse to overly complicated calculations.  
Based on the analysis of these three limits,  
we propose a possible schematic phase diagram, as shown in Fig.~\ref{PhaseDiagram}.

\subsection{Low-density expansion} \label{sec:low}
In this section, we consider the low-density regime defined by $k_F a_{\rm 2D} \ll 1$ and $k_F R_{\rm 2D} \ll 1$. In this 
limit, one can have trimer and dimer molecules in the ground state that are small compared to the typical distance between particles (which 
is set by $k_F^{-1}$). Therefore, we may approximate the state of the dilute system as a gas of interacting point-like fermionic trimers and/or bosonic dimers. 
Conservation of particle number implies that 
the number densities of dimers $n_d$ and trimers $n_t$ are related to the total atom density as follows:
\begin{equation}\label{number}
3 n= 3 n_t + 2 n_d \, .
\end{equation}
For a given density $n$, the ground state of the system corresponds to the set of densities $n_t$,  $n_d$ with the lowest energy per particle or, equivalently, the lowest chemical potential $\mu$. 

Let us start from the limit of vanishing density, $n \to 0$,  where we may neglect the intermolecular interactions. In this case, the trimer gas simply corresponds to a non-interacting Fermi gas, while the dimers form an ideal Bose-Einstein condensate.
Here, the chemical potentials of the trimer and dimer gases are, respectively, given by
\begin{align}\label{lowdensity}
\mu_t = \frac{E_3}{3} + \frac{2\pi n_t}{3m} \, , \hspace{6mm} & \mu_d = \frac{E_2}{2} \, ,
\end{align}
where $E_2$, $E_3$ are the few-body energies calculated in Sec.~\ref{three-body}, and 
 the second term in $\mu_t$ corresponds to the Fermi energy of the trimer gas. 
Comparing the different chemical potentials gives rise to three distinct regimes: (i) a trimer-only phase, where $n_d = 0$ and $\mu_t < \mu_d$; (ii) a trimer-dimer mixture, $\mu_t = \mu_d$; and (iii) a dimer-only phase, where $n_t = 0$ and $\mu_t > \mu_d$. These different phases are depicted in the low-density regime of the phase diagram in Fig.~\ref{PhaseDiagram}.

\begin{figure}
    \centering
    \includegraphics[width=0.9\columnwidth]{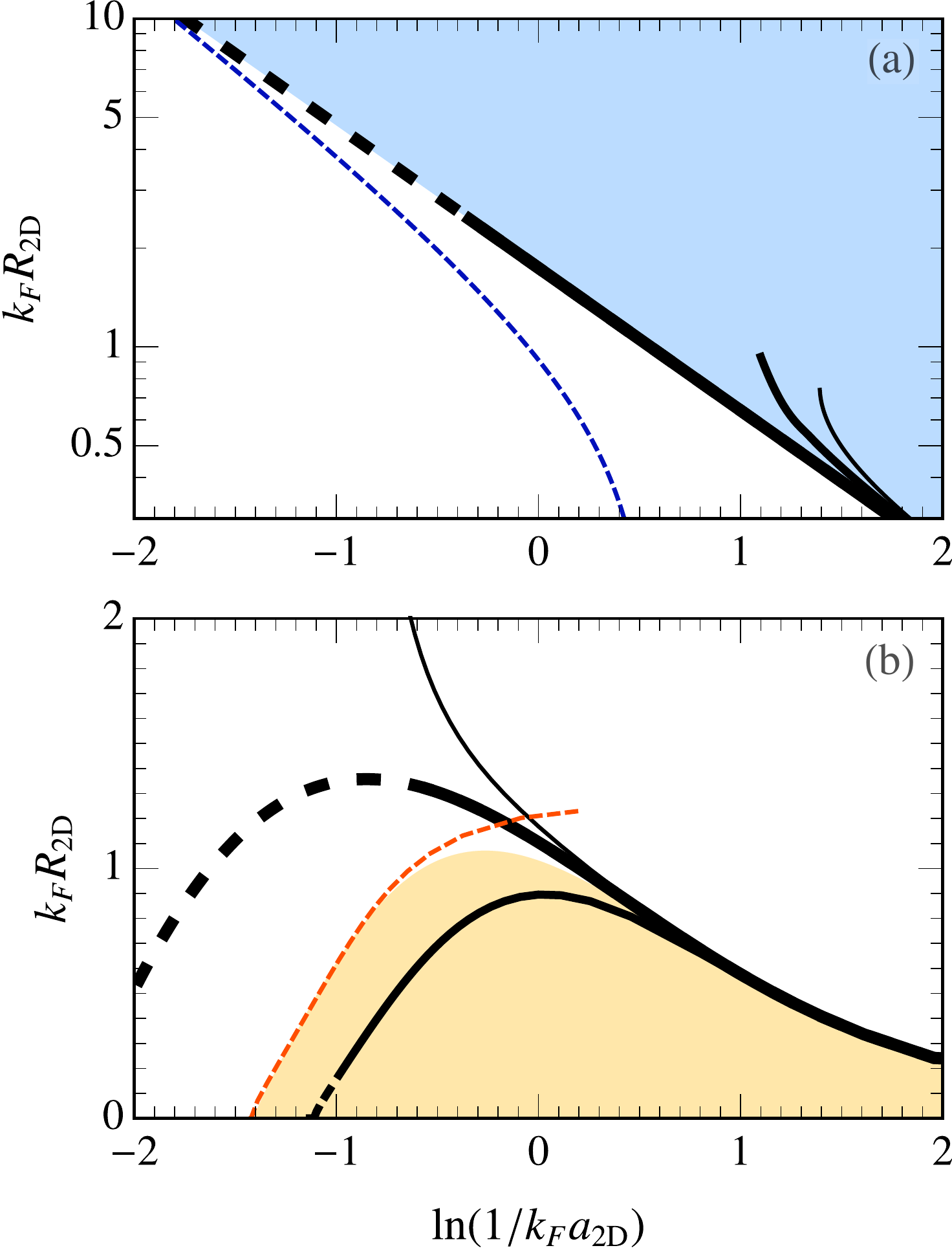}
    \caption{
    Phase boundaries for the dimer-only (a) and trimer-only (b) regions calculated within different approximations.
    The shaded regions are the same as in Fig.~\ref{PhaseDiagram}.
    (a) The solid (black) lines are defined by the low-density condition \eqref{notrimers}, where the thickest line corresponds to $a_{dd}=0$, while the thinner solid lines are for $a_{dd}=0.1 a_{\rm 2D}$ and $0.56 a_{\rm 2D}$, in order of decreasing thickness. The $a_{dd}=0$ phase boundary is displayed as dotted once the dimer size $\sim 1/\sqrt{m |E_2|}$ is larger than the interparticle spacing $\sim k_F^{-1}$.
    The thin dashed (blue) line corresponds to the mean-field phase boundary 
    described by Eq.~\eqref{MFT}. (b) The solid (black) lines depict the phase boundaries calculated using 
    Eq.~\eqref{nodimers}, where the thickest line is the solution with $\mathcal{F}=0$, while the thinner solid lines correspond to $\mathcal{F}=0.5$ and $-0.1$, in order of decreasing thickness.  
    The thick lines are displayed as dotted when the trimer size $\sim 1/\sqrt{m |E_3|}$ is larger than the interparticle spacing. The thin dashed (orange) line marks the region at high density where three-body correlations are relevant, as discussed  in Sec.~\ref{High-density}.
    This line is approximated by taking $\mathcal{F}=0.14$, which corresponds to the boundary of the (orange) shaded region.
  }\label{lowdensityplot}
\end{figure}

The phase boundary between the dimer-only and mixed phases is defined by the condition $E_2/2 = E_3/3$, i.e., we require the binding energy per atom to be equal for the dimers and trimers. As discussed in Sec.~\ref{three-body}, this yields  the curve $R_{\rm 2D}/a_{\rm 2D} \simeq 1.7$, 
which is shown as the thickest solid line in Fig.~\ref{lowdensityplot}(a).
Similarly, the boundary of the trimer-only region (where $n = n_t$) 
is described by the relation $E_2/2 = E_3/3 +2\pi n/3m$  and is shown as the thickest solid line in Fig.~\ref{lowdensityplot}(b). From this and Fig.~\ref{PhaseDiagram}, we see that the mixed trimer-dimer phase becomes a vanishingly small sliver of the phase diagram as we take $n \to 0$. This is a direct consequence of the fact that the trimer Fermi energy enters at a higher order in the density
than the molecular binding energies.

For larger densities, we must consider the intermolecular interactions  in the dilute gas, which can be described using the low-energy scattering amplitudes for the dominant partial-wave components. 
In particular, the trimer-trimer and dimer-dimer scattering amplitudes are,
respectively, given by:
\begin{align}
\begin{split}
\label{scatteringamplitudes}
  f_{tt}(k)  \simeq 4 s_{tt} k^2 \, , & \hspace{5mm}  
  f_{dd}(k)  \simeq \frac{4\pi}{\ln(\frac{1}{k^2a_{dd}^2})}\, .
\end{split}
\end{align}
Here, we have relative momentum $\mathbf{k}$ between the scattering particles in 2D, while $s_{tt}$ is the trimer-trimer scattering area, and $a_{dd}$ is the dimer-dimer scattering length.
At low energies, the bosonic nature of the dimer ensures that the dimer-dimer scattering is predominantly in the $s$-wave channel.
On the other hand, the low-energy expression for the trimer-trimer scattering amplitude has the general form of a $p$-wave interaction since the trimers are identical fermions, i.e., there is only one type of $s$-wave trimer.
Note that 
we have taken the angle between incoming and outgoing momenta in the scattering problem to be zero, so that $f_{tt}$ only depends on the magnitude $k = |\k|$.

In principle, there are three different flavors of dimer in the SU(3) fermion system, thus implying that we have both intraspecies and interspecies dimer-dimer scattering lengths, $a_{dd}$ and $a_{dd}^*$, respectively. However, since a trimer bound state is always present in the \textit{interspecies} dimer-dimer scattering problem (which involves three types of fermions rather than two), then we expect to have enhanced scattering such that $a^*_{dd} > a_{dd}$. Thus, a gas composed of identical dimers will have a lower energy than one containing three types of dimers. 
Furthermore, 
the identical-dimer ground state can still have equal densities for each fermion species $i$ if we pair within the transformed basis $i'$ defined by Eq.~\eqref{eq:transform}.

From the dimer-dimer scattering amplitude in \eqref{scatteringamplitudes}, we can incorporate dimer-dimer interactions into the chemical potential of the dimer gas as follows \cite{Q2DBackground_Review_Meera}:
\begin{align}
    \mu_d \simeq \frac{E_2}{2}+\frac{n_d}{2m}f_{dd}\left(\sqrt{4\pi n_d}\right),
\end{align}
where we have only kept the leading order term in $n_d a_{dd}^2$.
For the case of zero effective range, the dimer-dimer scattering length for two identical bosonic dimers  was determined to be $\left. a_{dd}\right|_{R_{\rm 2D}=0}=0.56 a_{\rm 2D}$~\cite{DimerDimer_Scattering_Petrov,DimerDimer_Scattering_Giorgini}. In general, the scattering length is unknown for $R_{\rm 2D} \neq 0$, but we expect $a_{dd}/a_{\rm 2D} \to 0$ as $R_{\rm 2D}/a_{\rm 2D} \to \infty$, since the system approaches a non-interacting Bose gas in this limit.
Since $a_{dd}$ appears in the argument of a 
logarithm, its precise value is not so important, provided that it is not exceptionally large, i.e., $a_{dd} \gg a_{\rm 2D}$, which is not expected to be the case here.

For the trimer gas, the chemical potential expanded up to lowest order in $n_t s_{tt}$ has the form
\begin{align*}
  \mu_t &\simeq \frac{E_3}{3}+\frac{2\pi n_t}{3m}+ \kappa
  \frac{n_t}{3m}f_{tt}\left(\sqrt{4\pi n_t}\right), 
\end{align*}
where $\kappa$ is a positive constant.
The precise determination of the scattering area $s_{tt}$ for two identical trimers is still an open problem. However, 
we can make progress without a  numerical value for $s_{tt}$ 
by using dimensional analysis to write: 
$\kappa s_{tt}=\mathcal{F}(R_{\rm 2D}/a_{\rm 2D}) a_{\rm 2D}^2$, where $\mathcal{F}$
is a dimensionless real function. This yields
\begin{align}
    \mu_t & \simeq \frac{E_3}{3}+\frac{2\pi n_t}{3m}\left(1 + 8 \mathcal{F} n_t a_{\rm 2D}^2 \right) .
\end{align}
We will discuss the possible behavior of $\mathcal{F}$ when we construct the phase diagram below.

In general, one must also consider the $s$-wave interactions between dimers and trimers when determining the phase diagram. 
Such trimer-dimer interactions can potentially result in phase separation between the trimer and dimer gases.
However, if we assume that the trimer-dimer interaction is sufficiently weak (i.e., the scattering length is sufficiently small) compared to the interactions in Eq.~\eqref{scatteringamplitudes}, then the phase boundaries between dimer-only, mixed and trimer-only phases should remain continuous. Therefore,
 the structure of the phase diagram  will be independent of the dimer-trimer interactions since either $n_t$ or $n_d$ will be zero on the phase boundaries.

Following the same procedure as in the non-interacting case, we can find the phase boundary between the trimer-only phase and the trimer-dimer mixture by setting $n_d=0$ and $\mu_t = \mu_d$. 
This yields the condition
\begin{equation} \label{nodimers}
\frac{E_2}{2}=\frac{E_3}{3}+\frac{2\pi n}{3m} \left(1 + 8 \mathcal{F} n a_{\rm 2D}^2 \right).
\end{equation}
Likewise, the boundary between the trimer-dimer mixture and the dimer-only phase is given by:
\begin{equation} \label{notrimers}
\frac{E_3}{3}=\frac{E_2}{2}+\frac{3n}{4m}f_{dd}(\sqrt{6\pi n}) ,
\end{equation}
where we have used the fact that $2n_d=3n$ in this case.

According to Eq.~\eqref{notrimers}, 
the presence of repulsive dimer-dimer interactions reduces the size of the dimer-only region in the phase diagram as the density is increased. 
To obtain an estimate of the phase boundary, we use the dimer-dimer scattering length $a_{dd}= 0.56 a_{\rm 2D}$, which is the known result for $R_{\rm 2D}=0$~\cite{DimerDimer_Scattering_Petrov,DimerDimer_Scattering_Giorgini}. Referring to Fig.~\ref{lowdensityplot}(a), we see that our estimated boundary substantially deviates from the non-interacting boundary (set by $E_2/2 = E_3/3$) once $k_F R_{\rm 2D} \approx 1$. However, at this point, $R_{\rm 2D}/a_{\rm 2D}$ approaches $\sim 10$, and thus we expect $a_{dd} \ll 0.56 a_{\rm 2D}$ since we have  $a_{dd}/a_{\rm 2D} \to 0$ in the limit $R_{\rm 2D}/a_{\rm 2D} \to \infty$, as discussed previously.
We therefore expect that the 
phase boundary initially follows the trajectory of the thinnest line in Fig.~\ref{lowdensityplot}(a), but eventually tends towards the non-interacting line as we increase $k_F R_{\rm 2D}$. This is represented schematically in the low-density regime of Fig.~\ref{PhaseDiagram}. Note that this transition remains continuous as long as the dimer-trimer interactions are irrelevant.

Turning to the phase boundary between the dimer-trimer mixture and the trimer-only phase, we see from Eq.~\eqref{nodimers} that
 the shape and size of the trimer-only region will depend on the behavior of $\mathcal{F}$. Though the trimer-trimer scattering problem is unsolved, we can deduce the likely magnitude and sign of $s_{tt}$, and therefore estimate $\mathcal{F}$.
In the absence of any scattering resonances, the strength of the trimer-trimer scattering should be set by $a_{\rm 2D}$ when $R_{\rm 2D} = 0$, which means that $\mathcal{F}$ should be of order 1 or less for arbitrary $R_{\rm 2D}$.

At first glance, one might conclude that the trimer-trimer \textit{p}-wave interactions are attractive, i.e., $\mathcal{F} < 0$, since there are no bound states with four or more particles in the six-body problem. For instance, in the simpler case of $p$-wave scattering between a 1-particle and a 1-2 dimer, the scattering area is always negative when the mass ratio $m_1/m_2 < 3.33$ and there are no
 three-body bound states \cite{pricoupenko2010,ngampruetikorn2013}. 
However, this atom-dimer scattering has a different symmetry to our trimer-trimer problem: The atom-dimer attraction is induced by an exchange process where the 2-particle can readily form a dimer with either of the 1-particles. This does not extend to the case of two trimers, since an atom in one of the trimers must be excited in order for an atom to be exchanged (see, for instance, the work on trimer-trimer scattering in the 3D two-component Fermi system~\cite{ResonatingGroupMethod_Shimpei}). It is therefore likely that the trimer-trimer \textit{p}-wave scattering is repulsive such that $\mathcal{F}>0$.

The boundary defined by Eq.~\eqref{nodimers} is shown in Fig.~\ref{lowdensityplot}(b) for attractive and repulsive trimer-trimer interactions. We see that the phase boundary becomes sensitive to $\mathcal{F}$ as the density increases, because, as one might expect, attractive interactions enlarge the trimer-only region, while repulsive interactions disfavor the trimer phase.
For the former case where $\mathcal{F}<0$, we expect the trimers to form a superfluid of \textit{p}-wave pairs, while for the latter case where $\mathcal{F}>0$, we will simply have a Fermi liquid of trimers  
\footnote{Strictly speaking, a repulsive Fermi gas will always form a superfluid at sufficiently low temperature, due to the effective interactions induced by the medium. However, since the critical temperature for superfluidity is exponentially suppressed compared to all other energy scales in the system,
we can consider the ground state to be a Fermi liquid}.
In the schematic phase diagram in Fig.~\ref{PhaseDiagram}, we have used repulsive trimer-trimer interactions, with $\mathcal{F} = 0.14$.

\subsection{BCS mean-field theory}
We now consider the limit where $k_F R_{\rm 2D}\gg 1$. 
This is the 2D analog of a narrow Feshbach resonance in 3D, where it is known that fluctuations around the mean-field approximation are suppressed by $1/k_F R_{\rm 3D}$~\cite{two-component_fermions_Gurarie,parish2007}. 
Since gaussian fluctuations around mean-field theory 
remain finite even in 2D~\cite{he2015quantum}, we expect such fluctuations to be similarly suppressed by $1/k_F R_{\rm 2D}$ as $R_{\rm 2D}\rightarrow \infty$.
Therefore, we will employ BCS mean-field theory to obtain the leading order behavior in $1/k_F R_{\rm 2D}$ in this limit.

In our three-component Fermi system, we require the density of all closed-channel boson flavours to be equal due to SU(3) symmetry. 
This implies that the condensate order parameters $\langle b_{\mathbf{0},j}\rangle$ are equal up to an arbitrary phase.
In the ground state, we may set the phases to zero, 
without loss of generality, and thus  
define a single, real, order parameter: $\Delta \equiv \sqrt{3}g|\langle b_{\mathbf{0},j} \rangle |$. 
Starting from Eq.~\eqref{hamiltonian}, the mean-field Hamiltonian is then given by~\cite{CondensedMatter_FieldTheory_Altland}:
\begin{equation} 
\hat{H}_{MF}
=\frac{1}{2}\sum_{\mathbf{k}} \Phi^{\dagger}_{\mathbf{k}} M_{\mathbf{k}} \Phi_{\mathbf{k}}+\frac{3}{2}\sum_{\mathbf{k}}\xi_\k+\frac{\Delta^2}{g^2}(\nu - 2\mu),
\end{equation}
where $\Phi_{\mathbf{k}} = (c_{\mathbf{k},1},c_{-\mathbf{k},1}^{\dagger},c_{\mathbf{k},2},c_{-\mathbf{k},2}^{\dagger}c_{\mathbf{k},3},c_{-\mathbf{k},3}^{\dagger})^T$,
\begin{equation}
M_{\mathbf{k}} =  
\begin{pmatrix}
 \xi_{\mathbf{k}} & 0 & 0 & \frac{\Delta}{\sqrt{3}} & 0 & \frac{\Delta}{\sqrt{3}} \\
  0 & -\xi_{\mathbf{k}} & -\frac{\Delta}{\sqrt{3}} & 0 & -\frac{\Delta}{\sqrt{3}} & 0 \\
  0  & -\frac{\Delta}{\sqrt{3}}  & \xi_{\mathbf{k}} & 0 & 0 & \frac{\Delta}{\sqrt{3}}  \\
  \frac{\Delta}{\sqrt{3}} & 0 & 0 & -\xi_{\mathbf{k}} & -\frac{\Delta}{\sqrt{3}} & 0 \\
  0 & -\frac{\Delta}{\sqrt{3}} & 0 & -\frac{\Delta}{\sqrt{3}} & \xi_{\mathbf{k}} &  0 \\
  \frac{\Delta}{\sqrt{3}} & 0 & \frac{\Delta}{\sqrt{3}} & 0 & 0 & -\xi_{\mathbf{k}}
 \end{pmatrix}\, ,
\end{equation}
and $\xi_{\mathbf{k}}=\epsilon_{\mathbf{k}}-\mu$. This matrix has six eigenvalues: $\pm \xi_\k$ and a degenerate pair $\pm E_{\k}$, where $E_{\k}=\sqrt{\xi_\k^2+\Delta^2}$. The former type of eigenvalue corresponds to an atomic Fermi sea, while the latter is the quasi-particle spectrum associated with BCS pairing.

At zero temperature, we have the mean-field free energy 
$\Omega =\langle\hat{H}_{MF}\rangle$, which is given by:
\begin{align} \label{MFTenergy}
\Omega =\sum_{\mathbf{k}}\xi_{\mathbf{k}}&\Theta\left(-\xi_\k\right)
+\sum_{\mathbf{k}}\left(\xi_{\mathbf{k}}-E_{\k}\right)
+\frac{\Delta^2}{g^2}(\nu - 2\mu)\, ,
\end{align}
where $\Theta(x)$ is the Heaviside step function.

Note that the mean-field approximation only incorporates pairing between fermions, and it does not allow for the possibility of trimers or three-body clustering. Minimizing the free energy gives rise to two possible phases: a BCS paired superfluid coexisting with an atomic Fermi sea for $\mu >0$, and a superfluid state where all the atoms are paired when $\mu < 0$. 
This scenario is consistent with previous BCS mean-field studies~\cite{Modawi1997,3-component_pairing_Paananen,3-component_FieldTheory_Symmetries_HighEnergy_He,3-component_BCSwavefunction_PhaseSeperation,3-component_finiteT_coexistence_Paananen,3-component_DomainWalls_Catelani,3-component_MediatedInteractions,3-component_BCSBECCrossover_Ozawa,Nummi2011,3-component_SpinOrbitCoupling_FFLO_Zhou}.

We calculate the boundary between these two phases by solving the gap and number equations, $\frac{\partial \Omega}{\partial \Delta}=0$ and $3n=-\frac{\partial \Omega}{\partial \mu}$, at $\mu=0$:
\begin{equation}\label{MFT}
k_F^2 a_{\rm 2D}^2 =\frac{R_{\rm 2D}^2}{3\pi a_{\rm 2D}^2}\exp\bigg(\frac{2R_{\rm 2D}^2}{a_{\rm 2D}^2}\bigg)+\frac{2}{3}\exp\bigg(\frac{R_{\rm 2D}^2}{a_{\rm 2D}^2}\bigg).
\end{equation}
This phase boundary is plotted as the dashed (blue) line in Fig.~\ref{lowdensityplot}(a), where the atom-pair mixture emerges as we decrease $\ln(1/k_Fa_{\rm 2D})$ at fixed $k_FR_{\rm 2D}$.

A natural question is how the phases in the limit $k_FR_{\rm 2D}\rightarrow\infty$ are related to the phases obtained in the low-density limit in Sec.~\ref{sec:low}. As discussed in Ref.~\cite{3-component_Background_Nishida} for the 3D case, one possibility is that the mean-field phase boundary smoothly connects with the boundary of the dimer-only phase at low density.
This scenario is represented schematically in Fig.~\ref{PhaseDiagram}, where we have the low-density dimer-trimer mixture smoothly evolving into the atom-pair phase with increasing $k_FR_{\rm 2D}$. Thus, we obtain an atom-trimer crossover, which is the cold-atom analog of the quark-hadron continuity in nuclear matter.

An alternative scenario is that there is an intervening phase or a first-order phase transition that destroys the crossover between atoms and trimers at intermediate densities. However, note that the 2D system is always stable against collapse (i.e., an unconstrained increase in the density), since the gas becomes weakly interacting with increasing density.

\subsection{High-density ansatz}\label{High-density}
To complete the phase diagram, we now tackle the high-density, weak-coupling regime $k_F a_{\rm 2D} \gg 1$. 
In the high-density limit $k_F a_{\rm 2D} \rightarrow \infty$, the ground state is the non-interacting gas, whose 
wave function is a product state of three distinguishable Fermi seas: $|FS\rangle = \prod_{|\mathbf{k}|\leq k_F,i}c_{\mathbf{k},i}^{\dagger}|0\rangle$. 
Therefore, to analyze the behavior near this limit, we consider perturbations of the non-interacting ground state that can accommodate different few-body correlations. 

Let us begin by looking at two-body correlations. Perturbing away from the high-density limit, the simplest 
correction to the ideal Fermi gas that includes two-body correlations is a wave function of the form:
\begin{align}\label{cooperpair}
  |\Psi^C_2\rangle &= \alpha b^{\dagger}_{\mathbf{0},1}|FS\rangle +\sum_{\mathbf{k}}\beta_{\mathbf{k}}c^{\dagger}_{\mathbf{k},2}c^{\dagger}_{-\mathbf{k},3}|FS\rangle .
\end{align}
Here, we have allowed  a single 2-3 pair of fermions to interact, while the rest of the fermions in the system remain non-interacting. Despite the lack of explicit interactions, the inert Fermi seas can effectively influence the correlations between the 2-3 pair via the Pauli exclusion principle. The problem of a single interacting pair of fermions placed above an ideal Fermi sea is equivalent to the Cooper pair problem~\cite{CooperPair_Original_Cooper} which was first introduced in the context of superconductivity, and which demonstrated that pairing in the presence of a Fermi sea exists for arbitrarily weak attraction. 

To determine the energy of the Cooper pair state $|\Psi_2^C\rangle$, we use the variational method, where 
we minimize the quantity $\langle\Psi_2^C|\hat{H}|\Psi_2^C\rangle$ over the set of variational parameters $\alpha$ and $\beta_{\k}$. This gives us an upper bound on the ground-state energy, which corresponds to the Cooper pair energy
\begin{align}
E_2^C &= 2\epsilon_F-\frac{1}{mR_{\rm 2D}^2}W\left( \frac{R_{\rm 2D}^2}{a_{\rm 2D}^2}  e^{k_F^2 R_{\rm 2D}^2}\right)\, ,
\end{align}
where we have defined the Fermi energy $\epsilon_F=k_F^2/2m$. We find that $E_2^C<2\epsilon_F$ for arbitrarily large $k_F a_{\rm 2D}$, which implies that the pairing of fermions is favored in the weak-coupling high-density regime. Note that the threshold for pairing has been shifted from zero in the vacuum case to $2\epsilon_F$ in the high-density case. In particular, for $R_{\rm 2D} =  0$, we simply  obtain $E_2^C =2\epsilon_F-1/ma_{\rm 2D}^2$, where $-1/ma_{\rm 2D}^2$ is the dimer energy at zero density.

Using the same concepts, we can write down a high-density ansatz for a \textit{Cooper trimer}: 
\begin{align*}
|& \Psi_3^C  \rangle = \sum_{\mathbf{k},i}\alpha_{\mathbf{k},i} b^{\dagger}_{\mathbf{k},i} c^{\dagger}_{-\mathbf{k},i}|FS\rangle \\ 
&+\sum_{\mathbf{k}_1,\mathbf{k}_2,\mathbf{k}_3}\beta_{\mathbf{k}_1 \mathbf{k}_2 \mathbf{k}_3}\, \delta (\mathbf{k}_1+\mathbf{k}_2+\mathbf{k}_3) \, c_{\mathbf{k}_1,1}^{\dagger}c_{\mathbf{k}_2,2}^{\dagger}c_{\mathbf{k}_3,3}^{\dagger}|FS\rangle \, .
\end{align*}
This corresponds to three distinguishable fermions interacting on top of the Fermi sea, thus allowing for three-body correlations. Once again, we
 proceed to minimize the quantity $\langle\Psi_3^C|\hat{H}|\Psi_3^C\rangle$ over the variational parameters $\alpha_{\k,i}$ and $\beta_{\k_1,\k_2,\k_3}$. The Cooper trimer energy satisfies an equation similar to that of the vacuum trimer as follows,
\begin{align}\notag
&\left[\frac{E_3^C-\nu-  \frac{3}{2}\epsilon_{\mathbf{k}}}{g^2}
-\sum_{|\mathbf{k}'|>k_F}\frac{\Theta(\epsilon_{\mathbf{k}+\mathbf{k}'}-\epsilon_{F})}{E^C_3-\epsilon_{\mathbf{k}}-\epsilon_{\mathbf{k}'}-\epsilon_{\mathbf{k}+\mathbf{k}'}} \right]
C_{\mathbf{k}}
\\
&\hspace{8mm}\notag =2\sum_{|\mathbf{k}'|>k_F}\frac{\Theta(\epsilon_{\mathbf{k}+\mathbf{k}'}-\epsilon_{F})C_{\mathbf{k}'}}{E^C_3-\epsilon_{\mathbf{k}}-\epsilon_{\mathbf{k}'}-\epsilon_{\mathbf{k}+\mathbf{k}'}} \, ,
\end{align}
where $|\k| > k_F$ and $C_\k = \sum_i \alpha_{\k,i}$.

Unlike the Cooper pair, we find that the Cooper trimer does not exist for arbitrary $k_F a_{\rm 2D}$. Instead, the three atoms prefer to remain uncorrelated, i.e., $E_3^C = 3\epsilon_F$, above a critical $k_Fa_{\rm 2D}$.
For $R_{\rm 2D}=0$, the Cooper trimer first appears at $\ln(1/k_Fa_{\rm 2D})\approx-1.47$. 
A similar situation is observed in 3D~\cite{niemann2012}, where for a given density, three distinguishable fermions require a sufficiently strong attraction in order to form a three-body cluster.  

In order to determine the preference of the atoms towards two- or three-body correlations, we compare the energy of the Cooper trimer $E_3^C$ with the energy of the Cooper pair plus one non-interacting atom: $E_2^C+\epsilon_F$. As discussed, the Cooper pair exists for all densities whereas the Cooper trimer only exists below a critical density; therefore, at sufficiently large density, the atoms will prefer to form two-body correlations rather than three-body correlations. On the other hand, in the low-density limit, we know that the three atoms prefer to form bound trimer states rather than bound dimer states, for sufficiently small $R_{\rm 2D}/a_{\rm 2D}$. Therefore, the condition $E_3^C=E_2^C+\epsilon_F$ defines a boundary between two-body and three-body correlations in the atomic Fermi gas. 
For $R_{\rm 2D}=0$, the transition is located at $\ln(1/a_{\rm 2D}k_F)\approx-1.42$, while the boundary for general $k_F R_{\rm 2D}$ is shown in Fig.~\ref{lowdensityplot}(b).

From Fig.~\ref{lowdensityplot}(b), we identify the correspondence between the results of our low-density expansion and the high-density ansatz. Specifically, we see that the low-density $\mathcal{F}=0.14$ curve interpolates between the two limits and we therefore use this curve for the schematic phase diagram in Fig.~\ref{PhaseDiagram}. We conjecture that the trimer-only phase of the low-density regime smoothly connects with the Cooper trimer phase of the high-density regime. Here, the two approximations conspire to form a dome-shaped region in the phase diagram for $k_F R_{\rm 2D} \lesssim 1.5$. Within this dome, the ground state is defined by strong three-body correlations which evolve from tightly bound trimers in the low-density regime to Cooper trimers in the high-density regime. In parallel to this evolution, 
above the dome in the phase diagram (Fig.~\ref{PhaseDiagram}),
the low-density trimer states evolve into an uncorrelated Fermi sea of atoms.

\section{Conclusion and outlook} \label{sec:conc}
In this work, we have investigated the few- and many-body behavior of a quasi-2D three-component Fermi gas with SU(3) symmetry.
We have focused on the regime of strong quasi-2D confinement,
where the system can be parametrized using the 2D scattering length and the 2D effective range. 
In the 2D limit, we have argued that the trimer state is expected to be longer lived compared with its 3D counterpart since it has a reduced weight at short distances and, consequently, three-body loss processes are suppressed. 
Moreover, we have demonstrated that the 2D trimer is always bound in the three-body system for arbitrary values of $R_{\rm 2D}/a_{\rm 2D}$, in contrast to the case in 3D.
These results all imply that trimers play an important role in the 2D three-component Fermi gas.
 
For the many-body system, we have constructed the phase diagram of the SU(3) Fermi gas by analyzing perturbations to the low- and high-density limits where the ground state is known. 
Our calculations suggest that trimers in the low-density limit can evolve into strong three-body correlations with increasing particle density. However, two-body correlations dominate in the limits where $k_F R_{\rm 2D} \gg 1$ or $\ln(1/k_F a_{\rm 2D}) \ll -1$.

Unlike the 3D case, the 2D Fermi gas becomes weakly interacting in the high-density limit, and it is thus expected to be stable against an unconstrained increase in the density, i.e., a collapse. Therefore, the 2D system may be more favorable for realising analogs of the quark-hadron continuity in nuclear matter, where fermionic quasiparticles smoothly change their character from atom-like to trimer-like with increasing attraction~\cite{rapp2007,3-component_Background_Nishida,3-component_Background2_Nishida}.
However, it remains an open question whether or not a first-order phase transition will disrupt such an atom-trimer crossover. As a first step towards addressing this problem, one would require a detailed analysis of the  trimer-trimer and trimer-dimer scattering amplitudes, similar to what has been done for $p$-wave trimers in 3D~\cite{ResonatingGroupMethod_Shimpei}.

The proposed phase diagram in Fig.~\ref{PhaseDiagram} can be investigated experimentally using $^6$Li atoms confined to a quasi-2D geometry. Since there are overlapping Feshbach resonances between the three lowest hyperfine states of $^6$Li, one can engineer a gas that is close to being SU(3) symmetric \cite{ottenstein2008collisional}. 
Even if we relax the SU(3) symmetry, we do not expect the overall phase structure of Fig.~\ref{PhaseDiagram} to change. In this case, the mixed atom-pair phase in the weak-coupling regime will now involve Cooper pairs comprised of the particles with the strongest attraction, while the fully paired superfluid phase will consist of a mixture of dimers that could potentially undergo phase separation.  Alternatively, one can realise a 2D SU(3) Fermi gas with $^{173}$Yb atoms~\cite{gorshkov2010two}, where the existence of orbital Feshbach resonances provides an interesting twist to the problem~\cite{PhysRevLett.115.135301,PhysRevLett.115.265302,pagano2015}.
In both cases, one can use radio-frequency pulses to directly probe or associate trimers in the three-component Fermi system~\cite{3-component_BindingEnergyMeasurement_Ueda,FequencyAssociation_EfimovTriemrs_jochim}.
Thus, to facilitate the experimental realization, 
we require a precise calculation of the quasi-2D three-body problem for realistic experimental parameters, which is the subject of future work.

\acknowledgments 
We are grateful to P. Crowley, S. Endo, A. Green,  J. Levinsen, Z. Shi, and L. Yeoh for fruitful discussions, and we thank F. Campaioli for help with figures.
MMP acknowledges support from the EPSRC under Grant No.\ EP/H00369X/2, and from the Australian Research Council via Discovery Project No.~DP160102739.

\appendix

\section{Decay rate derivation}\label{decayratederivation}
Consider an initial state $|\Psi (0)\rangle$ subjected to the evolution operator $\hat{O}(t)$, such that the state at time $t$ is given by $|\Psi(t)\rangle=\hat{O}(t)|\Psi(0)\rangle$. 
For the time-independent  Hamiltonian $\hat{H}$ in Eq.~\eqref{hamiltonian}, we have operator $\hat{O}(t)=e^{-i\hat{H}t}$ and we expect the amplitude $A(t)\equiv \langle \Psi(t)|\Psi(t)\rangle$ to be conserved at all times.

Now consider including the anti-Hermitian perturbation $\hat{H}_{3b}$ from Eq.~\eqref{perturbation}. In this case,
we have $\hat{O}(t)=e^{-i(\hat{H}+\hat{H}_{3b})t}$ and $\hat{O}^\dagger(t)=e^{i(\hat{H}-\hat{H}_{3b})t}$, so that the amplitude decreases in time as follows:
\begin{align*}
\dot{A}(t)&=\frac{d}{d t}\langle \Psi(t)|\Psi(t)\rangle \\
& = \bra{\Psi(0)} \left(\frac{d\hat{O}^\dag}{dt} \hat{O}(t) + \hat{O}^\dag(t) \frac{d\hat{O}}{dt} \right) \ket{\Psi(0)} 
\\
&=-2i\langle \Psi(t)|\hat{H}_{3b} |\Psi(t)\rangle \, .
\end{align*}
Since $\hat{H}_{3b}$ is a small perturbation, we can take $|\Psi(0)\rangle$ to be an eigenstate of $\hat{H}$.
Therefore, we can define the  decay rate $\Gamma = -\dot{A}(0)/A(0)$.
Using the three-body state in Eq.~\eqref{3wave}  and assuming $A(0) =1$, the necessary expectation value is therefore 
\begin{align}
\langle \Psi_3|\hat{H}_{3b}|\Psi_3\rangle =-\frac{i}{2}\frac{\Delta}{3}\Big|\sum_{\mathbf{k}}C_{\mathbf{k}}\Big|^2
\end{align}
and the decay rate is: 
\begin{align}\notag
\Gamma &= 2i\langle \Psi_3|\hat{H}_{3b}|\Psi_3\rangle\\
&=\frac{\Delta}{3}\Big|\sum_{\mathbf{k}}C_{\mathbf{k}}\Big|^2 \, ,
\end{align}
which is the result quoted in the main text.

\section{Spatial size of the trimer}\label{sizeappendix}
We wish to construct a convenient real-space three-body wave function in the SU(3) Fermi system from which we can estimate the size of the ground-state trimer. 
Working in the centre-of-mass frame, 
we start by defining the real-space
coordinates $\mathbf{s}_{ij}$ and $\mathbf{r}_{ij}$, where 
$\mathbf{s}_{ij}$ is the separation between a pair of particles $i$ and $j$; and $\mathbf{r}_{ij}$ is the distance from the centre-of-mass of the $i$-$j$ pair to the third particle.
To reduce it to a problem with only one coordinate, we take the separation between two particles to be zero. To be concrete, we take the pair of particles at zero separation to be 1 and 2 in the following.

Our three-body wave function in Eq.~\eqref{3wave} has two types of terms:  
an atom plus a closed-channel bosonic dimer, and a configuration involving
three atoms. 
For the atom-dimer (ad) parts of the wave function,
the closed-channel dimers are point-like and thus $\mathbf{s}_{ij}=0$ for any pair $i$, $j$ that makes up a closed-channel dimer. 
Since we have also set the separation of particles 1 and 2 to zero, we have
$\mathbf{r}_{31}=\mathbf{r}_{23}=0$. Defining $\mathbf{r}_{12}\equiv\mathbf{r}$, the transformed atom-dimer wave function is thus:
\begin{align}
\psi_{ad}(\mathbf{r})&=\sum_{\mathbf{k}} e^{i\mathbf{k}.\mathbf{r}
}\alpha_{\mathbf{k},3}
+ \sum_{\mathbf{k}}\left(\alpha_{\mathbf{k},2}
+\alpha_{\mathbf{k},1} \right)
\end{align}
where the second two terms just give a constant offset that is independent of $\mathbf{r}$.

Next we transform the three-atom (3a) component of the wave function, again focusing on the state where the  
$1$ and $2$ particles are located at the same point in space:
\begin{align}\notag
\psi_{3a}(\mathbf{r})= & \sum_{\mathbf{k}_1,\mathbf{k}_2,\mathbf{k}_3}\bigg[e^{i\mathbf{k}_3.\mathbf{r}}e^{i\frac{1}{2}(\mathbf{k}_1-\mathbf{k}_2).\mathbf{s}_{12}}\\
\notag
& \left. \times \beta_{\mathbf{k}_1 \mathbf{k}_2 \mathbf{k}_3} \, \delta (\mathbf{k}_1+\mathbf{k}_2+\mathbf{k}_3)\bigg]\right|_{\mathbf{s}_{12}=0
}
\\
&=\sum_{\mathbf{k}_3}e^{i\mathbf{k}_3.\mathbf{r}}\frac{1}{g}\eta_{\mathbf{k}_3}^{(3)} .
\end{align}
Here we have used the functions defined in Eq.~\eqref{eq:eta}. 
 From the relation $\left(E_3-\left(\frac {3}{2}\epsilon_{\mathbf{k}}+\nu\right)\right) \alpha_{\mathbf{k},i}   =  \eta_{\mathbf{k}}^{(i)}$, we see that
 $\psi_{ad}(\mathbf{r})/\psi_{3a}(\mathbf{r}) \to 0$
 when we take the limit $\Lambda \to \infty$ and $\nu\rightarrow \infty$. Therefore, we can neglect the atom-dimer contribution and
 define the normalized three-body wave function $\psi(\mathbf{r})$ as follows:
\begin{align*}
\psi(\mathbf{r})&= \lim_{\nu \to \infty} \frac{-\sum_{\mathbf{k}_3}e^{i\mathbf{k}_3.\mathbf{r}}\frac{1}{g} \eta_{\mathbf{k},3}}{\sqrt{\bigintss
\left|\sum_{\mathbf{k}_3}e^{i\mathbf{k}_3.\mathbf{r}}\frac{1}{g} \eta_{\mathbf{k},3}\right|^2d\mathbf{r}}}\\
& = \frac{\sum_{\mathbf{k}_3}e^{i\mathbf{k}_3.\mathbf{r}} \alpha_{\mathbf{k},3}}{\sqrt{\bigintss
\left|\sum_{\mathbf{k}_3}e^{i\mathbf{k}_3.\mathbf{r}} \alpha_{\mathbf{k},3}\right|^2d\mathbf{r}}}\\
&=\mathcal{N}^{-1/2}\sum_{\mathbf{k}} e^{i\mathbf{k}.\mathbf{r}}C_{\mathbf{k}} \, ,
\end{align*}
where we have used the relation $\alpha_{\mathbf{k},i}=\frac{1}{3}C_{\mathbf{k}}$ and defined the normalization factor $\mathcal{N}=\int  \left|\sum_{\mathbf{k}} e^{i\mathbf{k}.\mathbf{r}}C_{\mathbf{k}}\right|^2d\mathbf{r}$. We can now readily take the expectation value of the distance $r\equiv|\mathbf{r}|$,
as stated in Eq.~\eqref{eq:expect} of the main text.

\bibliography{3-component_Bibliography_v2_noDOI.bib}

\begin{thebibliography}{77}%
\makeatletter
\providecommand \@ifxundefined [1]{%
 \@ifx{#1\undefined}
}%
\providecommand \@ifnum [1]{%
 \ifnum #1\expandafter \@firstoftwo
 \else \expandafter \@secondoftwo
 \fi
}%
\providecommand \@ifx [1]{%
 \ifx #1\expandafter \@firstoftwo
 \else \expandafter \@secondoftwo
 \fi
}%
\providecommand \natexlab [1]{#1}%
\providecommand \enquote  [1]{``#1''}%
\providecommand \bibnamefont  [1]{#1}%
\providecommand \bibfnamefont [1]{#1}%
\providecommand \citenamefont [1]{#1}%
\providecommand \href@noop [0]{\@secondoftwo}%
\providecommand \href [0]{\begingroup \@sanitize@url \@href}%
\providecommand \@href[1]{\@@startlink{#1}\@@href}%
\providecommand \@@href[1]{\endgroup#1\@@endlink}%
\providecommand \@sanitize@url [0]{\catcode `\\12\catcode `\$12\catcode
  `\&12\catcode `\#12\catcode `\^12\catcode `\_12\catcode `\%12\relax}%
\providecommand \@@startlink[1]{}%
\providecommand \@@endlink[0]{}%
\providecommand \url  [0]{\begingroup\@sanitize@url \@url }%
\providecommand \@url [1]{\endgroup\@href {#1}{\urlprefix }}%
\providecommand \urlprefix  [0]{URL }%
\providecommand \Eprint [0]{\href }%
\providecommand \doibase [0]{http://dx.doi.org/}%
\providecommand \selectlanguage [0]{\@gobble}%
\providecommand \bibinfo  [0]{\@secondoftwo}%
\providecommand \bibfield  [0]{\@secondoftwo}%
\providecommand \translation [1]{[#1]}%
\providecommand \BibitemOpen [0]{}%
\providecommand \bibitemStop [0]{}%
\providecommand \bibitemNoStop [0]{.\EOS\space}%
\providecommand \EOS [0]{\spacefactor3000\relax}%
\providecommand \BibitemShut  [1]{\csname bibitem#1\endcsname}%
\let\auto@bib@innerbib\@empty
\bibitem [{\citenamefont {Rapp}\ \emph {et~al.}(2007)\citenamefont {Rapp},
  \citenamefont {Zar\'and}, \citenamefont {Honerkamp},\ and\ \citenamefont
  {Hofstetter}}]{rapp2007}%
  \BibitemOpen
  \bibfield  {author} {\bibinfo {author} {\bibfnamefont {A.}~\bibnamefont
  {Rapp}}, \bibinfo {author} {\bibfnamefont {G.}~\bibnamefont {Zar\'and}},
  \bibinfo {author} {\bibfnamefont {C.}~\bibnamefont {Honerkamp}}, \ and\
  \bibinfo {author} {\bibfnamefont {W.}~\bibnamefont {Hofstetter}},\ }\href
  {https://journals.aps.org/prl/abstract/10.1103/PhysRevLett.98.160405}
  {\bibfield  {journal} {\bibinfo  {journal} {Phys. Rev. Lett.}\ }\textbf
  {\bibinfo {volume} {98}},\ \bibinfo {pages} {160405} (\bibinfo {year}
  {2007})}\BibitemShut {NoStop}%
\bibitem [{\citenamefont {Nishida}(2012)}]{3-component_Background_Nishida}%
  \BibitemOpen
  \bibfield  {author} {\bibinfo {author} {\bibfnamefont {Y.}~\bibnamefont
  {Nishida}},\ }\href {http://link.aps.org/doi/10.1103/PhysRevLett.109.240401}
  {\bibfield  {journal} {\bibinfo  {journal} {Phys. Rev. Lett.}\ }\textbf
  {\bibinfo {volume} {109}},\ \bibinfo {pages} {240401} (\bibinfo {year}
  {2012})}\BibitemShut {NoStop}%
\bibitem [{\citenamefont {Nishida}(2015)}]{3-component_Background2_Nishida}%
  \BibitemOpen
  \bibfield  {author} {\bibinfo {author} {\bibfnamefont {Y.}~\bibnamefont
  {Nishida}},\ }\href {http://link.aps.org/doi/10.1103/PhysRevLett.114.115302}
  {\bibfield  {journal} {\bibinfo  {journal} {Phys. Rev. Lett.}\ }\textbf
  {\bibinfo {volume} {114}},\ \bibinfo {pages} {115302} (\bibinfo {year}
  {2015})}\BibitemShut {NoStop}%
\bibitem [{\citenamefont {Ottenstein}\ \emph {et~al.}(2008)\citenamefont
  {Ottenstein}, \citenamefont {Lompe}, \citenamefont {Kohnen}, \citenamefont
  {Wenz},\ and\ \citenamefont {Jochim}}]{ottenstein2008collisional}%
  \BibitemOpen
  \bibfield  {author} {\bibinfo {author} {\bibfnamefont {T.~B.}\ \bibnamefont
  {Ottenstein}}, \bibinfo {author} {\bibfnamefont {T.}~\bibnamefont {Lompe}},
  \bibinfo {author} {\bibfnamefont {M.}~\bibnamefont {Kohnen}}, \bibinfo
  {author} {\bibfnamefont {A.}~\bibnamefont {Wenz}}, \ and\ \bibinfo {author}
  {\bibfnamefont {S.}~\bibnamefont {Jochim}},\ }\href
  {https://journals.aps.org/prl/abstract/10.1103/PhysRevLett.101.203202}
  {\bibfield  {journal} {\bibinfo  {journal} {Phys. Rev. Lett.}\ }\textbf
  {\bibinfo {volume} {101}},\ \bibinfo {pages} {203202} (\bibinfo {year}
  {2008})}\BibitemShut {NoStop}%
\bibitem [{\citenamefont {Lompe}\ \emph
  {et~al.}(2010{\natexlab{a}})\citenamefont {Lompe}, \citenamefont
  {Ottenstein}, \citenamefont {Serwane}, \citenamefont {Wenz}, \citenamefont
  {Z{\"u}rn},\ and\ \citenamefont
  {Jochim}}]{FequencyAssociation_EfimovTriemrs_jochim}%
  \BibitemOpen
  \bibfield  {author} {\bibinfo {author} {\bibfnamefont {T.}~\bibnamefont
  {Lompe}}, \bibinfo {author} {\bibfnamefont {T.~B.}\ \bibnamefont
  {Ottenstein}}, \bibinfo {author} {\bibfnamefont {F.}~\bibnamefont {Serwane}},
  \bibinfo {author} {\bibfnamefont {A.~N.}\ \bibnamefont {Wenz}}, \bibinfo
  {author} {\bibfnamefont {G.}~\bibnamefont {Z{\"u}rn}}, \ and\ \bibinfo
  {author} {\bibfnamefont {S.}~\bibnamefont {Jochim}},\ }\href
  {http://science.sciencemag.org/content/330/6006/940} {\bibfield  {journal}
  {\bibinfo  {journal} {Science}\ }\textbf {\bibinfo {volume} {330}},\ \bibinfo
  {pages} {940} (\bibinfo {year} {2010}{\natexlab{a}})}\BibitemShut {NoStop}%
\bibitem [{\citenamefont {Nakajima}\ \emph {et~al.}(2011)\citenamefont
  {Nakajima}, \citenamefont {Horikoshi}, \citenamefont {Mukaiyama},
  \citenamefont {Naidon},\ and\ \citenamefont
  {Ueda}}]{3-component_BindingEnergyMeasurement_Ueda}%
  \BibitemOpen
  \bibfield  {author} {\bibinfo {author} {\bibfnamefont {S.}~\bibnamefont
  {Nakajima}}, \bibinfo {author} {\bibfnamefont {M.}~\bibnamefont {Horikoshi}},
  \bibinfo {author} {\bibfnamefont {T.}~\bibnamefont {Mukaiyama}}, \bibinfo
  {author} {\bibfnamefont {P.}~\bibnamefont {Naidon}}, \ and\ \bibinfo {author}
  {\bibfnamefont {M.}~\bibnamefont {Ueda}},\ }\href
  {http://link.aps.org/doi/10.1103/PhysRevLett.106.143201} {\bibfield
  {journal} {\bibinfo  {journal} {Phys. Rev. Lett.}\ }\textbf {\bibinfo
  {volume} {106}},\ \bibinfo {pages} {143201} (\bibinfo {year}
  {2011})}\BibitemShut {NoStop}%
\bibitem [{\citenamefont {Wenz}\ \emph {et~al.}(2009)\citenamefont {Wenz},
  \citenamefont {Lompe}, \citenamefont {Ottenstein}, \citenamefont {Serwane},
  \citenamefont {Z\"urn},\ and\ \citenamefont
  {Jochim}}]{3-component_EfimovLoss_Jochim}%
  \BibitemOpen
  \bibfield  {author} {\bibinfo {author} {\bibfnamefont {A.~N.}\ \bibnamefont
  {Wenz}}, \bibinfo {author} {\bibfnamefont {T.}~\bibnamefont {Lompe}},
  \bibinfo {author} {\bibfnamefont {T.~B.}\ \bibnamefont {Ottenstein}},
  \bibinfo {author} {\bibfnamefont {F.}~\bibnamefont {Serwane}}, \bibinfo
  {author} {\bibfnamefont {G.}~\bibnamefont {Z\"urn}}, \ and\ \bibinfo {author}
  {\bibfnamefont {S.}~\bibnamefont {Jochim}},\ }\href
  {http://link.aps.org/doi/10.1103/PhysRevA.80.040702} {\bibfield  {journal}
  {\bibinfo  {journal} {Phys. Rev. A}\ }\textbf {\bibinfo {volume} {80}},\
  \bibinfo {pages} {040702} (\bibinfo {year} {2009})}\BibitemShut {NoStop}%
\bibitem [{\citenamefont {Huckans}\ \emph {et~al.}(2009)\citenamefont
  {Huckans}, \citenamefont {Williams}, \citenamefont {Hazlett}, \citenamefont
  {Stites},\ and\ \citenamefont {O'Hara}}]{3-component_RecombinationLoss_Hara}%
  \BibitemOpen
  \bibfield  {author} {\bibinfo {author} {\bibfnamefont {J.~H.}\ \bibnamefont
  {Huckans}}, \bibinfo {author} {\bibfnamefont {J.~R.}\ \bibnamefont
  {Williams}}, \bibinfo {author} {\bibfnamefont {E.~L.}\ \bibnamefont
  {Hazlett}}, \bibinfo {author} {\bibfnamefont {R.~W.}\ \bibnamefont {Stites}},
  \ and\ \bibinfo {author} {\bibfnamefont {K.~M.}\ \bibnamefont {O'Hara}},\
  }\href {http://link.aps.org/doi/10.1103/PhysRevLett.102.165302} {\bibfield
  {journal} {\bibinfo  {journal} {Phys. Rev. Lett.}\ }\textbf {\bibinfo
  {volume} {102}},\ \bibinfo {pages} {165302} (\bibinfo {year}
  {2009})}\BibitemShut {NoStop}%
\bibitem [{\citenamefont {Williams}\ \emph {et~al.}(2009)\citenamefont
  {Williams}, \citenamefont {Hazlett}, \citenamefont {Huckans}, \citenamefont
  {Stites}, \citenamefont {Zhang},\ and\ \citenamefont
  {O'Hara}}]{3-component_Efimov_RecombinationLoss2_Hara}%
  \BibitemOpen
  \bibfield  {author} {\bibinfo {author} {\bibfnamefont {J.~R.}\ \bibnamefont
  {Williams}}, \bibinfo {author} {\bibfnamefont {E.~L.}\ \bibnamefont
  {Hazlett}}, \bibinfo {author} {\bibfnamefont {J.~H.}\ \bibnamefont
  {Huckans}}, \bibinfo {author} {\bibfnamefont {R.~W.}\ \bibnamefont {Stites}},
  \bibinfo {author} {\bibfnamefont {Y.}~\bibnamefont {Zhang}}, \ and\ \bibinfo
  {author} {\bibfnamefont {K.~M.}\ \bibnamefont {O'Hara}},\ }\href
  {http://link.aps.org/doi/10.1103/PhysRevLett.103.130404} {\bibfield
  {journal} {\bibinfo  {journal} {Phys. Rev. Lett.}\ }\textbf {\bibinfo
  {volume} {103}},\ \bibinfo {pages} {130404} (\bibinfo {year}
  {2009})}\BibitemShut {NoStop}%
\bibitem [{\citenamefont {Lompe}\ \emph
  {et~al.}(2010{\natexlab{b}})\citenamefont {Lompe}, \citenamefont
  {Ottenstein}, \citenamefont {Serwane}, \citenamefont {Viering}, \citenamefont
  {Wenz}, \citenamefont {Z\"urn},\ and\ \citenamefont
  {Jochim}}]{Efimov_AtomDimer_Resonance_Jochim}%
  \BibitemOpen
  \bibfield  {author} {\bibinfo {author} {\bibfnamefont {T.}~\bibnamefont
  {Lompe}}, \bibinfo {author} {\bibfnamefont {T.~B.}\ \bibnamefont
  {Ottenstein}}, \bibinfo {author} {\bibfnamefont {F.}~\bibnamefont {Serwane}},
  \bibinfo {author} {\bibfnamefont {K.}~\bibnamefont {Viering}}, \bibinfo
  {author} {\bibfnamefont {A.~N.}\ \bibnamefont {Wenz}}, \bibinfo {author}
  {\bibfnamefont {G.}~\bibnamefont {Z\"urn}}, \ and\ \bibinfo {author}
  {\bibfnamefont {S.}~\bibnamefont {Jochim}},\ }\href
  {http://link.aps.org/doi/10.1103/PhysRevLett.105.103201} {\bibfield
  {journal} {\bibinfo  {journal} {Phys. Rev. Lett.}\ }\textbf {\bibinfo
  {volume} {105}},\ \bibinfo {pages} {103201} (\bibinfo {year}
  {2010}{\natexlab{b}})}\BibitemShut {NoStop}%
\bibitem [{\citenamefont {Nakajima}\ \emph {et~al.}(2010)\citenamefont
  {Nakajima}, \citenamefont {Horikoshi}, \citenamefont {Mukaiyama},
  \citenamefont {Naidon},\ and\ \citenamefont
  {Ueda}}]{Efimov_AtomDimer_Resonance_Ueda}%
  \BibitemOpen
  \bibfield  {author} {\bibinfo {author} {\bibfnamefont {S.}~\bibnamefont
  {Nakajima}}, \bibinfo {author} {\bibfnamefont {M.}~\bibnamefont {Horikoshi}},
  \bibinfo {author} {\bibfnamefont {T.}~\bibnamefont {Mukaiyama}}, \bibinfo
  {author} {\bibfnamefont {P.}~\bibnamefont {Naidon}}, \ and\ \bibinfo {author}
  {\bibfnamefont {M.}~\bibnamefont {Ueda}},\ }\href
  {http://link.aps.org/doi/10.1103/PhysRevLett.105.023201} {\bibfield
  {journal} {\bibinfo  {journal} {Phys. Rev. Lett.}\ }\textbf {\bibinfo
  {volume} {105}},\ \bibinfo {pages} {023201} (\bibinfo {year}
  {2010})}\BibitemShut {NoStop}%
\bibitem [{\citenamefont {Efimov}(1970)}]{EfimovEffect_Efimov}%
  \BibitemOpen
  \bibfield  {author} {\bibinfo {author} {\bibfnamefont {V.}~\bibnamefont
  {Efimov}},\ }\href
  {http://www.sciencedirect.com/science/article/pii/0370269370903497}
  {\bibfield  {journal} {\bibinfo  {journal} {Physics Letters B}\ }\textbf
  {\bibinfo {volume} {33}},\ \bibinfo {pages} {563 } (\bibinfo {year}
  {1970})}\BibitemShut {NoStop}%
\bibitem [{\citenamefont {Braaten}\ and\ \citenamefont
  {Hammer}(2006)}]{ScatteringConcepts_NaturalLength_Universality_Efimov_Brateen}%
  \BibitemOpen
  \bibfield  {author} {\bibinfo {author} {\bibfnamefont {E.}~\bibnamefont
  {Braaten}}\ and\ \bibinfo {author} {\bibfnamefont {H.-W.}\ \bibnamefont
  {Hammer}},\ }\href
  {http://www.sciencedirect.com/science/article/pii/S0370157306000822}
  {\bibfield  {journal} {\bibinfo  {journal} {Physics Reports}\ }\textbf
  {\bibinfo {volume} {428}},\ \bibinfo {pages} {259 } (\bibinfo {year}
  {2006})}\BibitemShut {NoStop}%
\bibitem [{\citenamefont {Kraemer}\ \emph {et~al.}(2006)\citenamefont
  {Kraemer}, \citenamefont {Mark}, \citenamefont {Waldburger}, \citenamefont
  {Danzl}, \citenamefont {Chin}, \citenamefont {Engeser} \emph
  {et~al.}}]{EfimovObservation_Original_Kraemer}%
  \BibitemOpen
  \bibfield  {author} {\bibinfo {author} {\bibfnamefont {T.}~\bibnamefont
  {Kraemer}}, \bibinfo {author} {\bibfnamefont {M.}~\bibnamefont {Mark}},
  \bibinfo {author} {\bibfnamefont {P.}~\bibnamefont {Waldburger}}, \bibinfo
  {author} {\bibfnamefont {J.}~\bibnamefont {Danzl}}, \bibinfo {author}
  {\bibfnamefont {C.}~\bibnamefont {Chin}}, \bibinfo {author} {\bibnamefont
  {Engeser}},  \emph {et~al.},\ }\href
  {http://www.nature.com/nature/journal/v440/n7082/abs/nature04626.html}
  {\bibfield  {journal} {\bibinfo  {journal} {Nature}\ }\textbf {\bibinfo
  {volume} {440}},\ \bibinfo {pages} {315} (\bibinfo {year}
  {2006})}\BibitemShut {NoStop}%
\bibitem [{\citenamefont {Modawi}\ and\ \citenamefont
  {Leggett}(1997)}]{Modawi1997}%
  \BibitemOpen
  \bibfield  {author} {\bibinfo {author} {\bibfnamefont {A.~G.~K.}\
  \bibnamefont {Modawi}}\ and\ \bibinfo {author} {\bibfnamefont {A.~J.}\
  \bibnamefont {Leggett}},\ }\href {https://doi.org/10.1007/s10909-005-0103-3}
  {\bibfield  {journal} {\bibinfo  {journal} {Journal of Low Temperature
  Physics}\ }\textbf {\bibinfo {volume} {109}},\ \bibinfo {pages} {625}
  (\bibinfo {year} {1997})}\BibitemShut {NoStop}%
\bibitem [{\citenamefont {Paananen}\ \emph {et~al.}(2006)\citenamefont
  {Paananen}, \citenamefont {Martikainen},\ and\ \citenamefont
  {T\"orm\"a}}]{3-component_pairing_Paananen}%
  \BibitemOpen
  \bibfield  {author} {\bibinfo {author} {\bibfnamefont {T.}~\bibnamefont
  {Paananen}}, \bibinfo {author} {\bibfnamefont {J.-P.}\ \bibnamefont
  {Martikainen}}, \ and\ \bibinfo {author} {\bibfnamefont {P.}~\bibnamefont
  {T\"orm\"a}},\ }\href {http://link.aps.org/doi/10.1103/PhysRevA.73.053606}
  {\bibfield  {journal} {\bibinfo  {journal} {Phys. Rev. A}\ }\textbf {\bibinfo
  {volume} {73}},\ \bibinfo {pages} {053606} (\bibinfo {year}
  {2006})}\BibitemShut {NoStop}%
\bibitem [{\citenamefont {He}\ \emph {et~al.}(2006)\citenamefont {He},
  \citenamefont {Jin},\ and\ \citenamefont
  {Zhuang}}]{3-component_FieldTheory_Symmetries_HighEnergy_He}%
  \BibitemOpen
  \bibfield  {author} {\bibinfo {author} {\bibfnamefont {L.}~\bibnamefont
  {He}}, \bibinfo {author} {\bibfnamefont {M.}~\bibnamefont {Jin}}, \ and\
  \bibinfo {author} {\bibfnamefont {P.}~\bibnamefont {Zhuang}},\ }\href
  {http://link.aps.org/doi/10.1103/PhysRevA.74.033604} {\bibfield  {journal}
  {\bibinfo  {journal} {Phys. Rev. A}\ }\textbf {\bibinfo {volume} {74}},\
  \bibinfo {pages} {033604} (\bibinfo {year} {2006})}\BibitemShut {NoStop}%
\bibitem [{\citenamefont
  {Zhai}(2007)}]{3-component_BCSwavefunction_PhaseSeperation}%
  \BibitemOpen
  \bibfield  {author} {\bibinfo {author} {\bibfnamefont {H.}~\bibnamefont
  {Zhai}},\ }\href {http://link.aps.org/doi/10.1103/PhysRevA.75.031603}
  {\bibfield  {journal} {\bibinfo  {journal} {Phys. Rev. A}\ }\textbf {\bibinfo
  {volume} {75}},\ \bibinfo {pages} {031603} (\bibinfo {year}
  {2007})}\BibitemShut {NoStop}%
\bibitem [{\citenamefont {Paananen}\ \emph {et~al.}(2007)\citenamefont
  {Paananen}, \citenamefont {T\"orm\"a},\ and\ \citenamefont
  {Martikainen}}]{3-component_finiteT_coexistence_Paananen}%
  \BibitemOpen
  \bibfield  {author} {\bibinfo {author} {\bibfnamefont {T.}~\bibnamefont
  {Paananen}}, \bibinfo {author} {\bibfnamefont {P.}~\bibnamefont {T\"orm\"a}},
  \ and\ \bibinfo {author} {\bibfnamefont {J.-P.}\ \bibnamefont
  {Martikainen}},\ }\href {http://link.aps.org/doi/10.1103/PhysRevA.75.023622}
  {\bibfield  {journal} {\bibinfo  {journal} {Phys. Rev. A}\ }\textbf {\bibinfo
  {volume} {75}},\ \bibinfo {pages} {023622} (\bibinfo {year}
  {2007})}\BibitemShut {NoStop}%
\bibitem [{\citenamefont {Catelani}\ and\ \citenamefont
  {Yuzbashyan}(2008)}]{3-component_DomainWalls_Catelani}%
  \BibitemOpen
  \bibfield  {author} {\bibinfo {author} {\bibfnamefont {G.}~\bibnamefont
  {Catelani}}\ and\ \bibinfo {author} {\bibfnamefont {E.~A.}\ \bibnamefont
  {Yuzbashyan}},\ }\href {http://link.aps.org/doi/10.1103/PhysRevA.78.033615}
  {\bibfield  {journal} {\bibinfo  {journal} {Phys. Rev. A}\ }\textbf {\bibinfo
  {volume} {78}},\ \bibinfo {pages} {033615} (\bibinfo {year}
  {2008})}\BibitemShut {NoStop}%
\bibitem [{\citenamefont {Martikainen}\ \emph {et~al.}(2009)\citenamefont
  {Martikainen}, \citenamefont {Kinnunen}, \citenamefont {T\"orm\"a},\ and\
  \citenamefont {Pethick}}]{3-component_MediatedInteractions}%
  \BibitemOpen
  \bibfield  {author} {\bibinfo {author} {\bibfnamefont {J.-P.}\ \bibnamefont
  {Martikainen}}, \bibinfo {author} {\bibfnamefont {J.~J.}\ \bibnamefont
  {Kinnunen}}, \bibinfo {author} {\bibfnamefont {P.}~\bibnamefont {T\"orm\"a}},
  \ and\ \bibinfo {author} {\bibfnamefont {C.~J.}\ \bibnamefont {Pethick}},\
  }\href {http://link.aps.org/doi/10.1103/PhysRevLett.103.260403} {\bibfield
  {journal} {\bibinfo  {journal} {Phys. Rev. Lett.}\ }\textbf {\bibinfo
  {volume} {103}},\ \bibinfo {pages} {260403} (\bibinfo {year}
  {2009})}\BibitemShut {NoStop}%
\bibitem [{\citenamefont {Ozawa}\ and\ \citenamefont
  {Baym}(2010)}]{3-component_BCSBECCrossover_Ozawa}%
  \BibitemOpen
  \bibfield  {author} {\bibinfo {author} {\bibfnamefont {T.}~\bibnamefont
  {Ozawa}}\ and\ \bibinfo {author} {\bibfnamefont {G.}~\bibnamefont {Baym}},\
  }\href {http://link.aps.org/doi/10.1103/PhysRevA.82.063615} {\bibfield
  {journal} {\bibinfo  {journal} {Phys. Rev. A}\ }\textbf {\bibinfo {volume}
  {82}},\ \bibinfo {pages} {063615} (\bibinfo {year} {2010})}\BibitemShut
  {NoStop}%
\bibitem [{\citenamefont {Nummi}\ \emph {et~al.}(2011)\citenamefont {Nummi},
  \citenamefont {Kinnunen},\ and\ \citenamefont {T{\"o}rm{\"a}}}]{Nummi2011}%
  \BibitemOpen
  \bibfield  {author} {\bibinfo {author} {\bibfnamefont {O.~H.~T.}\
  \bibnamefont {Nummi}}, \bibinfo {author} {\bibfnamefont {J.~J.}\ \bibnamefont
  {Kinnunen}}, \ and\ \bibinfo {author} {\bibfnamefont {P.}~\bibnamefont
  {T{\"o}rm{\"a}}},\ }\href {http://stacks.iop.org/1367-2630/13/i=5/a=055013}
  {\bibfield  {journal} {\bibinfo  {journal} {New Journal of Physics}\ }\textbf
  {\bibinfo {volume} {13}},\ \bibinfo {pages} {055013} (\bibinfo {year}
  {2011})}\BibitemShut {NoStop}%
\bibitem [{\citenamefont {Zhou}\ \emph {et~al.}(2014)\citenamefont {Zhou},
  \citenamefont {Cui},\ and\ \citenamefont
  {Yi}}]{3-component_SpinOrbitCoupling_FFLO_Zhou}%
  \BibitemOpen
  \bibfield  {author} {\bibinfo {author} {\bibfnamefont {L.}~\bibnamefont
  {Zhou}}, \bibinfo {author} {\bibfnamefont {X.}~\bibnamefont {Cui}}, \ and\
  \bibinfo {author} {\bibfnamefont {W.}~\bibnamefont {Yi}},\ }\href
  {http://link.aps.org/doi/10.1103/PhysRevLett.112.195301} {\bibfield
  {journal} {\bibinfo  {journal} {Phys. Rev. Lett.}\ }\textbf {\bibinfo
  {volume} {112}},\ \bibinfo {pages} {195301} (\bibinfo {year}
  {2014})}\BibitemShut {NoStop}%
\bibitem [{\citenamefont {Floerchinger}\ \emph {et~al.}(2009)\citenamefont
  {Floerchinger}, \citenamefont {Schmidt}, \citenamefont {Moroz},\ and\
  \citenamefont {Wetterich}}]{3-component_RenomarlisationGroup_Floerchinger}%
  \BibitemOpen
  \bibfield  {author} {\bibinfo {author} {\bibfnamefont {S.}~\bibnamefont
  {Floerchinger}}, \bibinfo {author} {\bibfnamefont {R.}~\bibnamefont
  {Schmidt}}, \bibinfo {author} {\bibfnamefont {S.}~\bibnamefont {Moroz}}, \
  and\ \bibinfo {author} {\bibfnamefont {C.}~\bibnamefont {Wetterich}},\ }\href
  {http://link.aps.org/doi/10.1103/PhysRevA.79.013603} {\bibfield  {journal}
  {\bibinfo  {journal} {Phys. Rev. A}\ }\textbf {\bibinfo {volume} {79}},\
  \bibinfo {pages} {013603} (\bibinfo {year} {2009})}\BibitemShut {NoStop}%
\bibitem [{\citenamefont {Bedaque}\ and\ \citenamefont
  {D'Incao}(2009)}]{3-component_3D_unbalanceda_Incao}%
  \BibitemOpen
  \bibfield  {author} {\bibinfo {author} {\bibfnamefont {P.~F.}\ \bibnamefont
  {Bedaque}}\ and\ \bibinfo {author} {\bibfnamefont {J.~P.}\ \bibnamefont
  {D'Incao}},\ }\href
  {http://www.sciencedirect.com/science/article/pii/S0003491609000566}
  {\bibfield  {journal} {\bibinfo  {journal} {Annals of Physics}\ }\textbf
  {\bibinfo {volume} {324}},\ \bibinfo {pages} {1763 } (\bibinfo {year}
  {2009})}\BibitemShut {NoStop}%
\bibitem [{\citenamefont {Niemann}\ and\ \citenamefont
  {Hammer}(2012)}]{niemann2012}%
  \BibitemOpen
  \bibfield  {author} {\bibinfo {author} {\bibfnamefont {P.}~\bibnamefont
  {Niemann}}\ and\ \bibinfo {author} {\bibfnamefont {H.-W.}\ \bibnamefont
  {Hammer}},\ }\href {http://link.aps.org/doi/10.1103/PhysRevA.86.013628}
  {\bibfield  {journal} {\bibinfo  {journal} {Phys. Rev. A}\ }\textbf {\bibinfo
  {volume} {86}},\ \bibinfo {pages} {013628} (\bibinfo {year}
  {2012})}\BibitemShut {NoStop}%
\bibitem [{\citenamefont {Yi}\ and\ \citenamefont {Cui}(2015)}]{cui2015}%
  \BibitemOpen
  \bibfield  {author} {\bibinfo {author} {\bibfnamefont {W.}~\bibnamefont
  {Yi}}\ and\ \bibinfo {author} {\bibfnamefont {X.}~\bibnamefont {Cui}},\
  }\href {http://link.aps.org/doi/10.1103/PhysRevA.92.013620} {\bibfield
  {journal} {\bibinfo  {journal} {Phys. Rev. A}\ }\textbf {\bibinfo {volume}
  {92}},\ \bibinfo {pages} {013620} (\bibinfo {year} {2015})}\BibitemShut
  {NoStop}%
\bibitem [{\citenamefont {Azaria}\ \emph {et~al.}(2009)\citenamefont {Azaria},
  \citenamefont {Capponi},\ and\ \citenamefont {Lecheminant}}]{azaria2009}%
  \BibitemOpen
  \bibfield  {author} {\bibinfo {author} {\bibfnamefont {P.}~\bibnamefont
  {Azaria}}, \bibinfo {author} {\bibfnamefont {S.}~\bibnamefont {Capponi}}, \
  and\ \bibinfo {author} {\bibfnamefont {P.}~\bibnamefont {Lecheminant}},\
  }\href {https://link.aps.org/doi/10.1103/PhysRevA.80.041604} {\bibfield
  {journal} {\bibinfo  {journal} {Phys. Rev. A}\ }\textbf {\bibinfo {volume}
  {80}},\ \bibinfo {pages} {041604} (\bibinfo {year} {2009})}\BibitemShut
  {NoStop}%
\bibitem [{\citenamefont {Blume}\ \emph {et~al.}(2008)\citenamefont {Blume},
  \citenamefont {Rittenhouse}, \citenamefont {von Stecher},\ and\ \citenamefont
  {Greene}}]{Blume2008}%
  \BibitemOpen
  \bibfield  {author} {\bibinfo {author} {\bibfnamefont {D.}~\bibnamefont
  {Blume}}, \bibinfo {author} {\bibfnamefont {S.~T.}\ \bibnamefont
  {Rittenhouse}}, \bibinfo {author} {\bibfnamefont {J.}~\bibnamefont {von
  Stecher}}, \ and\ \bibinfo {author} {\bibfnamefont {C.~H.}\ \bibnamefont
  {Greene}},\ }\href {http://link.aps.org/doi/10.1103/PhysRevA.77.033627}
  {\bibfield  {journal} {\bibinfo  {journal} {Phys. Rev. A}\ }\textbf {\bibinfo
  {volume} {77}},\ \bibinfo {pages} {033627} (\bibinfo {year}
  {2008})}\BibitemShut {NoStop}%
\bibitem [{\citenamefont {Braaten}\ \emph {et~al.}(2010)\citenamefont
  {Braaten}, \citenamefont {Hammer}, \citenamefont {Kang},\ and\ \citenamefont
  {Platter}}]{Efimov_Lithium6_Theory_Hammer}%
  \BibitemOpen
  \bibfield  {author} {\bibinfo {author} {\bibfnamefont {E.}~\bibnamefont
  {Braaten}}, \bibinfo {author} {\bibfnamefont {H.~W.}\ \bibnamefont {Hammer}},
  \bibinfo {author} {\bibfnamefont {D.}~\bibnamefont {Kang}}, \ and\ \bibinfo
  {author} {\bibfnamefont {L.}~\bibnamefont {Platter}},\ }\href
  {http://link.aps.org/doi/10.1103/PhysRevA.81.013605} {\bibfield  {journal}
  {\bibinfo  {journal} {Phys. Rev. A}\ }\textbf {\bibinfo {volume} {81}},\
  \bibinfo {pages} {013605} (\bibinfo {year} {2010})}\BibitemShut {NoStop}%
\bibitem [{\citenamefont {Levinsen}\ \emph {et~al.}(2014)\citenamefont
  {Levinsen}, \citenamefont {Massignan},\ and\ \citenamefont
  {Parish}}]{Quasi2D_EfimovTrimers_Parish}%
  \BibitemOpen
  \bibfield  {author} {\bibinfo {author} {\bibfnamefont {J.}~\bibnamefont
  {Levinsen}}, \bibinfo {author} {\bibfnamefont {P.}~\bibnamefont {Massignan}},
  \ and\ \bibinfo {author} {\bibfnamefont {M.~M.}\ \bibnamefont {Parish}},\
  }\href {http://link.aps.org/doi/10.1103/PhysRevX.4.031020} {\bibfield
  {journal} {\bibinfo  {journal} {Phys. Rev. X}\ }\textbf {\bibinfo {volume}
  {4}},\ \bibinfo {pages} {031020} (\bibinfo {year} {2014})}\BibitemShut
  {NoStop}%
\bibitem [{\citenamefont {Yamashita}\ \emph {et~al.}(2015)\citenamefont
  {Yamashita}, \citenamefont {Bellotti}, \citenamefont {Frederico},
  \citenamefont {Fedorov}, \citenamefont {Jensen},\ and\ \citenamefont
  {Zinner}}]{yamashita2015}%
  \BibitemOpen
  \bibfield  {author} {\bibinfo {author} {\bibfnamefont {M.~T.}\ \bibnamefont
  {Yamashita}}, \bibinfo {author} {\bibfnamefont {F.~F.}\ \bibnamefont
  {Bellotti}}, \bibinfo {author} {\bibfnamefont {T.}~\bibnamefont {Frederico}},
  \bibinfo {author} {\bibfnamefont {D.~V.}\ \bibnamefont {Fedorov}}, \bibinfo
  {author} {\bibfnamefont {A.~S.}\ \bibnamefont {Jensen}}, \ and\ \bibinfo
  {author} {\bibfnamefont {N.~T.}\ \bibnamefont {Zinner}},\ }\href
  {http://stacks.iop.org/0953-4075/48/i=2/a=025302} {\bibfield  {journal}
  {\bibinfo  {journal} {Journal of Physics B: Atomic, Molecular and Optical
  Physics}\ }\textbf {\bibinfo {volume} {48}},\ \bibinfo {pages} {025302}
  (\bibinfo {year} {2015})}\BibitemShut {NoStop}%
\bibitem [{\citenamefont {D'Incao}\ \emph {et~al.}(2015)\citenamefont
  {D'Incao}, \citenamefont {Anis},\ and\ \citenamefont
  {Esry}}]{3BodyRecombination_2D__Incao}%
  \BibitemOpen
  \bibfield  {author} {\bibinfo {author} {\bibfnamefont {J.~P.}\ \bibnamefont
  {D'Incao}}, \bibinfo {author} {\bibfnamefont {F.}~\bibnamefont {Anis}}, \
  and\ \bibinfo {author} {\bibfnamefont {B.~D.}\ \bibnamefont {Esry}},\ }\href
  {http://link.aps.org/doi/10.1103/PhysRevA.91.062710} {\bibfield  {journal}
  {\bibinfo  {journal} {Phys. Rev. A}\ }\textbf {\bibinfo {volume} {91}},\
  \bibinfo {pages} {062710} (\bibinfo {year} {2015})}\BibitemShut {NoStop}%
\bibitem [{\citenamefont {Helfrich}\ and\ \citenamefont
  {Hammer}(2011)}]{ThreeBosons_2D_Hammer}%
  \BibitemOpen
  \bibfield  {author} {\bibinfo {author} {\bibfnamefont {K.}~\bibnamefont
  {Helfrich}}\ and\ \bibinfo {author} {\bibfnamefont {H.-W.}\ \bibnamefont
  {Hammer}},\ }\href {http://link.aps.org/doi/10.1103/PhysRevA.83.052703}
  {\bibfield  {journal} {\bibinfo  {journal} {Phys. Rev. A}\ }\textbf {\bibinfo
  {volume} {83}},\ \bibinfo {pages} {052703} (\bibinfo {year}
  {2011})}\BibitemShut {NoStop}%
\bibitem [{\citenamefont {Bruch}\ and\ \citenamefont
  {Tjon}(1979)}]{bruch1979binding}%
  \BibitemOpen
  \bibfield  {author} {\bibinfo {author} {\bibfnamefont {L.}~\bibnamefont
  {Bruch}}\ and\ \bibinfo {author} {\bibfnamefont {J.}~\bibnamefont {Tjon}},\
  }\href {http://journals.aps.org/pra/abstract/10.1103/PhysRevA.19.425}
  {\bibfield  {journal} {\bibinfo  {journal} {Phys. Rev. A}\ }\textbf {\bibinfo
  {volume} {19}},\ \bibinfo {pages} {425} (\bibinfo {year} {1979})}\BibitemShut
  {NoStop}%
\bibitem [{\citenamefont {G\"unter}\ \emph {et~al.}(2005)\citenamefont
  {G\"unter}, \citenamefont {St\"oferle}, \citenamefont {Moritz}, \citenamefont
  {K\"ohl},\ and\ \citenamefont
  {Esslinger}}]{2Dexperiments_pwavefermigas_Gunter}%
  \BibitemOpen
  \bibfield  {author} {\bibinfo {author} {\bibfnamefont {K.}~\bibnamefont
  {G\"unter}}, \bibinfo {author} {\bibfnamefont {T.}~\bibnamefont
  {St\"oferle}}, \bibinfo {author} {\bibfnamefont {H.}~\bibnamefont {Moritz}},
  \bibinfo {author} {\bibfnamefont {M.}~\bibnamefont {K\"ohl}}, \ and\ \bibinfo
  {author} {\bibfnamefont {T.}~\bibnamefont {Esslinger}},\ }\href
  {http://link.aps.org/doi/10.1103/PhysRevLett.95.230401} {\bibfield  {journal}
  {\bibinfo  {journal} {Phys. Rev. Lett.}\ }\textbf {\bibinfo {volume} {95}},\
  \bibinfo {pages} {230401} (\bibinfo {year} {2005})}\BibitemShut {NoStop}%
\bibitem [{\citenamefont {Martiyanov}\ \emph {et~al.}(2010)\citenamefont
  {Martiyanov}, \citenamefont {Makhalov},\ and\ \citenamefont
  {Turlapov}}]{2Dgasobservation_fermions_Turlapov}%
  \BibitemOpen
  \bibfield  {author} {\bibinfo {author} {\bibfnamefont {K.}~\bibnamefont
  {Martiyanov}}, \bibinfo {author} {\bibfnamefont {V.}~\bibnamefont
  {Makhalov}}, \ and\ \bibinfo {author} {\bibfnamefont {A.}~\bibnamefont
  {Turlapov}},\ }\href {http://link.aps.org/doi/10.1103/PhysRevLett.105.030404}
  {\bibfield  {journal} {\bibinfo  {journal} {Phys. Rev. Lett.}\ }\textbf
  {\bibinfo {volume} {105}},\ \bibinfo {pages} {030404} (\bibinfo {year}
  {2010})}\BibitemShut {NoStop}%
\bibitem [{\citenamefont {Fr\"ohlich}\ \emph {et~al.}(2011)\citenamefont
  {Fr\"ohlich}, \citenamefont {Feld}, \citenamefont {Vogt}, \citenamefont
  {Koschorreck}, \citenamefont {Zwerger},\ and\ \citenamefont
  {K\"ohl}}]{2Dexperiment_fermions_Zwerger}%
  \BibitemOpen
  \bibfield  {author} {\bibinfo {author} {\bibfnamefont {B.}~\bibnamefont
  {Fr\"ohlich}}, \bibinfo {author} {\bibfnamefont {M.}~\bibnamefont {Feld}},
  \bibinfo {author} {\bibfnamefont {E.}~\bibnamefont {Vogt}}, \bibinfo {author}
  {\bibfnamefont {M.}~\bibnamefont {Koschorreck}}, \bibinfo {author}
  {\bibfnamefont {W.}~\bibnamefont {Zwerger}}, \ and\ \bibinfo {author}
  {\bibfnamefont {M.}~\bibnamefont {K\"ohl}},\ }\href
  {http://link.aps.org/doi/10.1103/PhysRevLett.106.105301} {\bibfield
  {journal} {\bibinfo  {journal} {Phys. Rev. Lett.}\ }\textbf {\bibinfo
  {volume} {106}},\ \bibinfo {pages} {105301} (\bibinfo {year}
  {2011})}\BibitemShut {NoStop}%
\bibitem [{\citenamefont {Dyke}\ \emph {et~al.}(2011)\citenamefont {Dyke},
  \citenamefont {Kuhnle}, \citenamefont {Whitlock}, \citenamefont {Hu},
  \citenamefont {Mark}, \citenamefont {Hoinka}, \citenamefont {Lingham},
  \citenamefont {Hannaford},\ and\ \citenamefont
  {Vale}}]{From2Dto3D_Experimental_Vale}%
  \BibitemOpen
  \bibfield  {author} {\bibinfo {author} {\bibfnamefont {P.}~\bibnamefont
  {Dyke}}, \bibinfo {author} {\bibfnamefont {E.~D.}\ \bibnamefont {Kuhnle}},
  \bibinfo {author} {\bibfnamefont {S.}~\bibnamefont {Whitlock}}, \bibinfo
  {author} {\bibfnamefont {H.}~\bibnamefont {Hu}}, \bibinfo {author}
  {\bibfnamefont {M.}~\bibnamefont {Mark}}, \bibinfo {author} {\bibfnamefont
  {S.}~\bibnamefont {Hoinka}}, \bibinfo {author} {\bibfnamefont
  {M.}~\bibnamefont {Lingham}}, \bibinfo {author} {\bibfnamefont
  {P.}~\bibnamefont {Hannaford}}, \ and\ \bibinfo {author} {\bibfnamefont
  {C.~J.}\ \bibnamefont {Vale}},\ }\href
  {http://link.aps.org/doi/10.1103/PhysRevLett.106.105304} {\bibfield
  {journal} {\bibinfo  {journal} {Phys. Rev. Lett.}\ }\textbf {\bibinfo
  {volume} {106}},\ \bibinfo {pages} {105304} (\bibinfo {year}
  {2011})}\BibitemShut {NoStop}%
\bibitem [{\citenamefont {Boettcher}\ \emph {et~al.}(2016)\citenamefont
  {Boettcher}, \citenamefont {Bayha}, \citenamefont {Kedar}, \citenamefont
  {Murthy}, \citenamefont {Neidig}, \citenamefont {Ries}, \citenamefont {Wenz},
  \citenamefont {Z\"urn}, \citenamefont {Jochim},\ and\ \citenamefont
  {Enss}}]{3-component_reference_Experimental_Quasi2D_Pairing_Experimental_Jochim}%
  \BibitemOpen
  \bibfield  {author} {\bibinfo {author} {\bibfnamefont {I.}~\bibnamefont
  {Boettcher}}, \bibinfo {author} {\bibfnamefont {L.}~\bibnamefont {Bayha}},
  \bibinfo {author} {\bibfnamefont {D.}~\bibnamefont {Kedar}}, \bibinfo
  {author} {\bibfnamefont {P.~A.}\ \bibnamefont {Murthy}}, \bibinfo {author}
  {\bibfnamefont {M.}~\bibnamefont {Neidig}}, \bibinfo {author} {\bibfnamefont
  {M.~G.}\ \bibnamefont {Ries}}, \bibinfo {author} {\bibfnamefont {A.~N.}\
  \bibnamefont {Wenz}}, \bibinfo {author} {\bibfnamefont {G.}~\bibnamefont
  {Z\"urn}}, \bibinfo {author} {\bibfnamefont {S.}~\bibnamefont {Jochim}}, \
  and\ \bibinfo {author} {\bibfnamefont {T.}~\bibnamefont {Enss}},\ }\href
  {http://link.aps.org/doi/10.1103/PhysRevLett.116.045303} {\bibfield
  {journal} {\bibinfo  {journal} {Phys. Rev. Lett.}\ }\textbf {\bibinfo
  {volume} {116}},\ \bibinfo {pages} {045303} (\bibinfo {year}
  {2016})}\BibitemShut {NoStop}%
\bibitem [{\citenamefont {Feld}\ \emph {et~al.}(2011)\citenamefont {Feld},
  \citenamefont {Fr{\"o}hlich}, \citenamefont {Vogt}, \citenamefont
  {Koschorreck},\ and\ \citenamefont
  {K{\"o}hl}}]{2Dexperiment_pairingpseudogap_Kohl}%
  \BibitemOpen
  \bibfield  {author} {\bibinfo {author} {\bibfnamefont {M.}~\bibnamefont
  {Feld}}, \bibinfo {author} {\bibfnamefont {B.}~\bibnamefont {Fr{\"o}hlich}},
  \bibinfo {author} {\bibfnamefont {E.}~\bibnamefont {Vogt}}, \bibinfo {author}
  {\bibfnamefont {M.}~\bibnamefont {Koschorreck}}, \ and\ \bibinfo {author}
  {\bibfnamefont {M.}~\bibnamefont {K{\"o}hl}},\ }\href
  {http://www.nature.com/nature/journal/v480/n7375/full/nature10627.html}
  {\bibfield  {journal} {\bibinfo  {journal} {Nature}\ }\textbf {\bibinfo
  {volume} {480}},\ \bibinfo {pages} {75} (\bibinfo {year} {2011})}\BibitemShut
  {NoStop}%
\bibitem [{\citenamefont {Sommer}\ \emph {et~al.}(2012)\citenamefont {Sommer},
  \citenamefont {Cheuk}, \citenamefont {Ku}, \citenamefont {Bakr},\ and\
  \citenamefont
  {Zwierlein}}]{2Dexperiment_tightconfinement_Lithium6_Zwierlein}%
  \BibitemOpen
  \bibfield  {author} {\bibinfo {author} {\bibfnamefont {A.~T.}\ \bibnamefont
  {Sommer}}, \bibinfo {author} {\bibfnamefont {L.~W.}\ \bibnamefont {Cheuk}},
  \bibinfo {author} {\bibfnamefont {M.~J.~H.}\ \bibnamefont {Ku}}, \bibinfo
  {author} {\bibfnamefont {W.~S.}\ \bibnamefont {Bakr}}, \ and\ \bibinfo
  {author} {\bibfnamefont {M.~W.}\ \bibnamefont {Zwierlein}},\ }\href
  {http://link.aps.org/doi/10.1103/PhysRevLett.108.045302} {\bibfield
  {journal} {\bibinfo  {journal} {Phys. Rev. Lett.}\ }\textbf {\bibinfo
  {volume} {108}},\ \bibinfo {pages} {045302} (\bibinfo {year}
  {2012})}\BibitemShut {NoStop}%
\bibitem [{\citenamefont {Zhang}\ \emph {et~al.}(2012)\citenamefont {Zhang},
  \citenamefont {Ong}, \citenamefont {Arakelyan},\ and\ \citenamefont
  {Thomas}}]{2Dexperiment_polarons_Thomas}%
  \BibitemOpen
  \bibfield  {author} {\bibinfo {author} {\bibfnamefont {Y.}~\bibnamefont
  {Zhang}}, \bibinfo {author} {\bibfnamefont {W.}~\bibnamefont {Ong}}, \bibinfo
  {author} {\bibfnamefont {I.}~\bibnamefont {Arakelyan}}, \ and\ \bibinfo
  {author} {\bibfnamefont {J.~E.}\ \bibnamefont {Thomas}},\ }\href
  {http://link.aps.org/doi/10.1103/PhysRevLett.108.235302} {\bibfield
  {journal} {\bibinfo  {journal} {Phys. Rev. Lett.}\ }\textbf {\bibinfo
  {volume} {108}},\ \bibinfo {pages} {235302} (\bibinfo {year}
  {2012})}\BibitemShut {NoStop}%
\bibitem [{\citenamefont {Baur}\ \emph {et~al.}(2012)\citenamefont {Baur},
  \citenamefont {Fr\"ohlich}, \citenamefont {Feld}, \citenamefont {Vogt},
  \citenamefont {Pertot}, \citenamefont {Koschorreck},\ and\ \citenamefont
  {K\"ohl}}]{2Dexperiment_Feshbachmolecules_Kohl}%
  \BibitemOpen
  \bibfield  {author} {\bibinfo {author} {\bibfnamefont {S.~K.}\ \bibnamefont
  {Baur}}, \bibinfo {author} {\bibfnamefont {B.}~\bibnamefont {Fr\"ohlich}},
  \bibinfo {author} {\bibfnamefont {M.}~\bibnamefont {Feld}}, \bibinfo {author}
  {\bibfnamefont {E.}~\bibnamefont {Vogt}}, \bibinfo {author} {\bibfnamefont
  {D.}~\bibnamefont {Pertot}}, \bibinfo {author} {\bibfnamefont
  {M.}~\bibnamefont {Koschorreck}}, \ and\ \bibinfo {author} {\bibfnamefont
  {M.}~\bibnamefont {K\"ohl}},\ }\href
  {http://link.aps.org/doi/10.1103/PhysRevA.85.061604} {\bibfield  {journal}
  {\bibinfo  {journal} {Phys. Rev. A}\ }\textbf {\bibinfo {volume} {85}},\
  \bibinfo {pages} {061604} (\bibinfo {year} {2012})}\BibitemShut {NoStop}%
\bibitem [{\citenamefont {Koschorreck}\ \emph {et~al.}(2012)\citenamefont
  {Koschorreck}, \citenamefont {Pertot}, \citenamefont {Vogt}, \citenamefont
  {Fr{\"o}hlich}, \citenamefont {Feld},\ and\ \citenamefont
  {K{\"o}hl}}]{2Dexperiment_repulsivefermions}%
  \BibitemOpen
  \bibfield  {author} {\bibinfo {author} {\bibfnamefont {M.}~\bibnamefont
  {Koschorreck}}, \bibinfo {author} {\bibfnamefont {D.}~\bibnamefont {Pertot}},
  \bibinfo {author} {\bibfnamefont {E.}~\bibnamefont {Vogt}}, \bibinfo {author}
  {\bibfnamefont {B.}~\bibnamefont {Fr{\"o}hlich}}, \bibinfo {author}
  {\bibfnamefont {M.}~\bibnamefont {Feld}}, \ and\ \bibinfo {author}
  {\bibfnamefont {M.}~\bibnamefont {K{\"o}hl}},\ }\href
  {http://www.nature.com/nature/journal/v485/n7400/full/nature11151.html}
  {\bibfield  {journal} {\bibinfo  {journal} {Nature}\ }\textbf {\bibinfo
  {volume} {485}},\ \bibinfo {pages} {619} (\bibinfo {year}
  {2012})}\BibitemShut {NoStop}%
\bibitem [{\citenamefont {Vogt}\ \emph {et~al.}(2012)\citenamefont {Vogt},
  \citenamefont {Feld}, \citenamefont {Fr\"ohlich}, \citenamefont {Pertot},
  \citenamefont {Koschorreck},\ and\ \citenamefont
  {K\"ohl}}]{2Dexperiment_fermiliquids_kohl}%
  \BibitemOpen
  \bibfield  {author} {\bibinfo {author} {\bibfnamefont {E.}~\bibnamefont
  {Vogt}}, \bibinfo {author} {\bibfnamefont {M.}~\bibnamefont {Feld}}, \bibinfo
  {author} {\bibfnamefont {B.}~\bibnamefont {Fr\"ohlich}}, \bibinfo {author}
  {\bibfnamefont {D.}~\bibnamefont {Pertot}}, \bibinfo {author} {\bibfnamefont
  {M.}~\bibnamefont {Koschorreck}}, \ and\ \bibinfo {author} {\bibfnamefont
  {M.}~\bibnamefont {K\"ohl}},\ }\href
  {http://link.aps.org/doi/10.1103/PhysRevLett.108.070404} {\bibfield
  {journal} {\bibinfo  {journal} {Phys. Rev. Lett.}\ }\textbf {\bibinfo
  {volume} {108}},\ \bibinfo {pages} {070404} (\bibinfo {year}
  {2012})}\BibitemShut {NoStop}%
\bibitem [{\citenamefont {Makhalov}\ \emph {et~al.}(2014)\citenamefont
  {Makhalov}, \citenamefont {Martiyanov},\ and\ \citenamefont
  {Turlapov}}]{2Dexperiment_pressure_Turlapov}%
  \BibitemOpen
  \bibfield  {author} {\bibinfo {author} {\bibfnamefont {V.}~\bibnamefont
  {Makhalov}}, \bibinfo {author} {\bibfnamefont {K.}~\bibnamefont
  {Martiyanov}}, \ and\ \bibinfo {author} {\bibfnamefont {A.}~\bibnamefont
  {Turlapov}},\ }\href {http://link.aps.org/doi/10.1103/PhysRevLett.112.045301}
  {\bibfield  {journal} {\bibinfo  {journal} {Phys. Rev. Lett.}\ }\textbf
  {\bibinfo {volume} {112}},\ \bibinfo {pages} {045301} (\bibinfo {year}
  {2014})}\BibitemShut {NoStop}%
\bibitem [{\citenamefont {Qin}\ \emph {et~al.}(2015)\citenamefont {Qin},
  \citenamefont {Wu}, \citenamefont {Zhang}, \citenamefont {Yi},\ and\
  \citenamefont {Guo}}]{3-component_2D_FFLO_Chang}%
  \BibitemOpen
  \bibfield  {author} {\bibinfo {author} {\bibfnamefont {F.}~\bibnamefont
  {Qin}}, \bibinfo {author} {\bibfnamefont {F.}~\bibnamefont {Wu}}, \bibinfo
  {author} {\bibfnamefont {W.}~\bibnamefont {Zhang}}, \bibinfo {author}
  {\bibfnamefont {W.}~\bibnamefont {Yi}}, \ and\ \bibinfo {author}
  {\bibfnamefont {G.-C.}\ \bibnamefont {Guo}},\ }\href
  {http://link.aps.org/doi/10.1103/PhysRevA.92.023604} {\bibfield  {journal}
  {\bibinfo  {journal} {Phys. Rev. A}\ }\textbf {\bibinfo {volume} {92}},\
  \bibinfo {pages} {023604} (\bibinfo {year} {2015})}\BibitemShut {NoStop}%
\bibitem [{\citenamefont {Chin}\ \emph {et~al.}(2010)\citenamefont {Chin},
  \citenamefont {Grimm}, \citenamefont {Julienne},\ and\ \citenamefont
  {Tiesinga}}]{FeshbachResonanca_Review_Grimm}%
  \BibitemOpen
  \bibfield  {author} {\bibinfo {author} {\bibfnamefont {C.}~\bibnamefont
  {Chin}}, \bibinfo {author} {\bibfnamefont {R.}~\bibnamefont {Grimm}},
  \bibinfo {author} {\bibfnamefont {P.}~\bibnamefont {Julienne}}, \ and\
  \bibinfo {author} {\bibfnamefont {E.}~\bibnamefont {Tiesinga}},\ }\href
  {http://link.aps.org/doi/10.1103/RevModPhys.82.1225} {\bibfield  {journal}
  {\bibinfo  {journal} {Rev. Mod. Phys.}\ }\textbf {\bibinfo {volume} {82}},\
  \bibinfo {pages} {1225} (\bibinfo {year} {2010})}\BibitemShut {NoStop}%
\bibitem [{\citenamefont {Bartenstein}\ \emph {et~al.}(2005)\citenamefont
  {Bartenstein}, \citenamefont {Altmeyer}, \citenamefont {Riedl}, \citenamefont
  {Geursen}, \citenamefont {Jochim}, \citenamefont {Chin}, \citenamefont
  {Denschlag}, \citenamefont {Grimm}, \citenamefont {Simoni}, \citenamefont
  {Tiesinga}, \citenamefont {Williams},\ and\ \citenamefont
  {Julienne}}]{3-component_reference_Experimental_FeshbachRsonancesLithium_ScatteringLength_Bartenstein}%
  \BibitemOpen
  \bibfield  {author} {\bibinfo {author} {\bibfnamefont {M.}~\bibnamefont
  {Bartenstein}}, \bibinfo {author} {\bibfnamefont {A.}~\bibnamefont
  {Altmeyer}}, \bibinfo {author} {\bibfnamefont {S.}~\bibnamefont {Riedl}},
  \bibinfo {author} {\bibfnamefont {R.}~\bibnamefont {Geursen}}, \bibinfo
  {author} {\bibfnamefont {S.}~\bibnamefont {Jochim}}, \bibinfo {author}
  {\bibfnamefont {C.}~\bibnamefont {Chin}}, \bibinfo {author} {\bibfnamefont
  {J.~H.}\ \bibnamefont {Denschlag}}, \bibinfo {author} {\bibfnamefont
  {R.}~\bibnamefont {Grimm}}, \bibinfo {author} {\bibfnamefont
  {A.}~\bibnamefont {Simoni}}, \bibinfo {author} {\bibfnamefont
  {E.}~\bibnamefont {Tiesinga}}, \bibinfo {author} {\bibfnamefont {C.~J.}\
  \bibnamefont {Williams}}, \ and\ \bibinfo {author} {\bibfnamefont {P.~S.}\
  \bibnamefont {Julienne}},\ }\href
  {http://link.aps.org/doi/10.1103/PhysRevLett.94.103201} {\bibfield  {journal}
  {\bibinfo  {journal} {Phys. Rev. Lett.}\ }\textbf {\bibinfo {volume} {94}},\
  \bibinfo {pages} {103201} (\bibinfo {year} {2005})}\BibitemShut {NoStop}%
\bibitem [{\citenamefont {Petrov}\ and\ \citenamefont
  {Shlyapnikov}(2001)}]{PhysRevA.64.012706}%
  \BibitemOpen
  \bibfield  {author} {\bibinfo {author} {\bibfnamefont {D.~S.}\ \bibnamefont
  {Petrov}}\ and\ \bibinfo {author} {\bibfnamefont {G.~V.}\ \bibnamefont
  {Shlyapnikov}},\ }\href {http://link.aps.org/doi/10.1103/PhysRevA.64.012706}
  {\bibfield  {journal} {\bibinfo  {journal} {Phys. Rev. A}\ }\textbf {\bibinfo
  {volume} {64}},\ \bibinfo {pages} {012706} (\bibinfo {year}
  {2001})}\BibitemShut {NoStop}%
\bibitem [{\citenamefont {Levinsen}\ and\ \citenamefont
  {Parish}(2015)}]{Q2DBackground_Review_Meera}%
  \BibitemOpen
  \bibfield  {author} {\bibinfo {author} {\bibfnamefont {J.}~\bibnamefont
  {Levinsen}}\ and\ \bibinfo {author} {\bibfnamefont {M.~M.}\ \bibnamefont
  {Parish}},\ }\href
  {http://www.worldscientific.com/worldscibooks/10.1142/9561} {\bibfield
  {journal} {\bibinfo  {journal} {Annual Review of Cold Atoms and Molecules}\
  }\textbf {\bibinfo {volume} {3}},\ \bibinfo {pages} {1} (\bibinfo {year}
  {2015})}\BibitemShut {NoStop}%
\bibitem [{\citenamefont {Adhikari}(1986)}]{2D_scatteringtheory_Adhikari}%
  \BibitemOpen
  \bibfield  {author} {\bibinfo {author} {\bibfnamefont {S.~K.}\ \bibnamefont
  {Adhikari}},\ }\href {\doibase http://dx.doi.org/10.1119/1.14623} {\bibfield
  {journal} {\bibinfo  {journal} {American Journal of Physics}\ }\textbf
  {\bibinfo {volume} {54}},\ \bibinfo {pages} {362} (\bibinfo {year}
  {1986})}\BibitemShut {NoStop}%
\bibitem [{\citenamefont {Gurarie}\ and\ \citenamefont
  {Radzihovsky}(2007)}]{two-component_fermions_Gurarie}%
  \BibitemOpen
  \bibfield  {author} {\bibinfo {author} {\bibfnamefont {V.}~\bibnamefont
  {Gurarie}}\ and\ \bibinfo {author} {\bibfnamefont {L.}~\bibnamefont
  {Radzihovsky}},\ }\href
  {http://www.sciencedirect.com/science/article/pii/S0003491606002399}
  {\bibfield  {journal} {\bibinfo  {journal} {Annals of Physics}\ }\textbf
  {\bibinfo {volume} {322}},\ \bibinfo {pages} {2 } (\bibinfo {year}
  {2007})}\BibitemShut {NoStop}%
\bibitem [{\citenamefont {Landau}\ and\ \citenamefont
  {Lifshitz}(1981)}]{landlandau}%
  \BibitemOpen
  \bibfield  {author} {\bibinfo {author} {\bibfnamefont {L.}~\bibnamefont
  {Landau}}\ and\ \bibinfo {author} {\bibfnamefont {E.}~\bibnamefont
  {Lifshitz}},\ }\href {https://books.google.co.uk/books?id=SvdoN3k8EysC}
  {\emph {\bibinfo {title} {Quantum Mechanics: Non-Relativistic Theory}}},\
  \bibinfo {series} {Course of Theoretical Physics}, Vol.~\bibinfo {volume}
  {3}\ (\bibinfo  {publisher} {Elsevier Science},\ \bibinfo {year}
  {1981})\BibitemShut {NoStop}%
\bibitem [{\citenamefont {Randeria}\ \emph {et~al.}(1989)\citenamefont
  {Randeria}, \citenamefont {Duan},\ and\ \citenamefont
  {Shieh}}]{PhysRevLett.62.981}%
  \BibitemOpen
  \bibfield  {author} {\bibinfo {author} {\bibfnamefont {M.}~\bibnamefont
  {Randeria}}, \bibinfo {author} {\bibfnamefont {J.-M.}\ \bibnamefont {Duan}},
  \ and\ \bibinfo {author} {\bibfnamefont {L.-Y.}\ \bibnamefont {Shieh}},\
  }\href {http://link.aps.org/doi/10.1103/PhysRevLett.62.981} {\bibfield
  {journal} {\bibinfo  {journal} {Phys. Rev. Lett.}\ }\textbf {\bibinfo
  {volume} {62}},\ \bibinfo {pages} {981} (\bibinfo {year} {1989})}\BibitemShut
  {NoStop}%
\bibitem [{\citenamefont {Levinsen}\ \emph {et~al.}(2009)\citenamefont
  {Levinsen}, \citenamefont {Tiecke}, \citenamefont {Walraven},\ and\
  \citenamefont {Petrov}}]{LongLiveTrimers_2-component_Jesper}%
  \BibitemOpen
  \bibfield  {author} {\bibinfo {author} {\bibfnamefont {J.}~\bibnamefont
  {Levinsen}}, \bibinfo {author} {\bibfnamefont {T.~G.}\ \bibnamefont
  {Tiecke}}, \bibinfo {author} {\bibfnamefont {J.~T.~M.}\ \bibnamefont
  {Walraven}}, \ and\ \bibinfo {author} {\bibfnamefont {D.~S.}\ \bibnamefont
  {Petrov}},\ }\href {http://link.aps.org/doi/10.1103/PhysRevLett.103.153202}
  {\bibfield  {journal} {\bibinfo  {journal} {Phys. Rev. Lett.}\ }\textbf
  {\bibinfo {volume} {103}},\ \bibinfo {pages} {153202} (\bibinfo {year}
  {2009})}\BibitemShut {NoStop}%
\bibitem [{\citenamefont {Naidon}\ and\ \citenamefont
  {Endo}(2017)}]{0034-4885-80-5-056001}%
  \BibitemOpen
  \bibfield  {author} {\bibinfo {author} {\bibfnamefont {P.}~\bibnamefont
  {Naidon}}\ and\ \bibinfo {author} {\bibfnamefont {S.}~\bibnamefont {Endo}},\
  }\href {http://stacks.iop.org/0034-4885/80/i=5/a=056001} {\bibfield
  {journal} {\bibinfo  {journal} {Reports on Progress in Physics}\ }\textbf
  {\bibinfo {volume} {80}},\ \bibinfo {pages} {056001} (\bibinfo {year}
  {2017})}\BibitemShut {NoStop}%
\bibitem [{\citenamefont {Kartavtsev}\ and\ \citenamefont
  {Malykh}(2007)}]{KartavtsevMalykh_Trimers_Kartavtsev}%
  \BibitemOpen
  \bibfield  {author} {\bibinfo {author} {\bibfnamefont {O.~I.}\ \bibnamefont
  {Kartavtsev}}\ and\ \bibinfo {author} {\bibfnamefont {A.~V.}\ \bibnamefont
  {Malykh}},\ }\href {http://stacks.iop.org/0953-4075/40/i=7/a=011} {\bibfield
  {journal} {\bibinfo  {journal} {Journal of Physics B: Atomic, Molecular and
  Optical Physics}\ }\textbf {\bibinfo {volume} {40}},\ \bibinfo {pages} {1429}
  (\bibinfo {year} {2007})}\BibitemShut {NoStop}%
\bibitem [{\citenamefont {Endo}\ \emph {et~al.}(2012)\citenamefont {Endo},
  \citenamefont {Naidon},\ and\ \citenamefont
  {Ueda}}]{Efimov_KartavtsevMalykh_CrossoverTrimers_Shimpei}%
  \BibitemOpen
  \bibfield  {author} {\bibinfo {author} {\bibfnamefont {S.}~\bibnamefont
  {Endo}}, \bibinfo {author} {\bibfnamefont {P.}~\bibnamefont {Naidon}}, \ and\
  \bibinfo {author} {\bibfnamefont {M.}~\bibnamefont {Ueda}},\ }\href
  {http://link.aps.org/doi/10.1103/PhysRevA.86.062703} {\bibfield  {journal}
  {\bibinfo  {journal} {Phys. Rev. A}\ }\textbf {\bibinfo {volume} {86}},\
  \bibinfo {pages} {062703} (\bibinfo {year} {2012})}\BibitemShut {NoStop}%
\bibitem [{\citenamefont {Levinsen}\ and\ \citenamefont
  {Petrov}(2011)}]{AtomDimer_DimerDimerScattering_Jesper}%
  \BibitemOpen
  \bibfield  {author} {\bibinfo {author} {\bibfnamefont {J.}~\bibnamefont
  {Levinsen}}\ and\ \bibinfo {author} {\bibfnamefont {D.~S.}\ \bibnamefont
  {Petrov}},\ }\href {http://dx.doi.org/10.1140/epjd/e2011-20071-x} {\bibfield
  {journal} {\bibinfo  {journal} {The European Physical Journal D}\ }\textbf
  {\bibinfo {volume} {65}},\ \bibinfo {pages} {67} (\bibinfo {year}
  {2011})}\BibitemShut {NoStop}%
\bibitem [{\citenamefont {Jag}\ \emph {et~al.}(2016)\citenamefont {Jag},
  \citenamefont {Cetina}, \citenamefont {Lous}, \citenamefont {Grimm},
  \citenamefont {Levinsen},\ and\ \citenamefont {Petrov}}]{jag2016}%
  \BibitemOpen
  \bibfield  {author} {\bibinfo {author} {\bibfnamefont {M.}~\bibnamefont
  {Jag}}, \bibinfo {author} {\bibfnamefont {M.}~\bibnamefont {Cetina}},
  \bibinfo {author} {\bibfnamefont {R.~S.}\ \bibnamefont {Lous}}, \bibinfo
  {author} {\bibfnamefont {R.}~\bibnamefont {Grimm}}, \bibinfo {author}
  {\bibfnamefont {J.}~\bibnamefont {Levinsen}}, \ and\ \bibinfo {author}
  {\bibfnamefont {D.~S.}\ \bibnamefont {Petrov}},\ }\href
  {https://link.aps.org/doi/10.1103/PhysRevA.94.062706} {\bibfield  {journal}
  {\bibinfo  {journal} {Phys. Rev. A}\ }\textbf {\bibinfo {volume} {94}},\
  \bibinfo {pages} {062706} (\bibinfo {year} {2016})}\BibitemShut {NoStop}%
\bibitem [{\citenamefont {Petrov}\ \emph {et~al.}(2003)\citenamefont {Petrov},
  \citenamefont {Baranov},\ and\ \citenamefont
  {Shlyapnikov}}]{DimerDimer_Scattering_Petrov}%
  \BibitemOpen
  \bibfield  {author} {\bibinfo {author} {\bibfnamefont {D.~S.}\ \bibnamefont
  {Petrov}}, \bibinfo {author} {\bibfnamefont {M.~A.}\ \bibnamefont {Baranov}},
  \ and\ \bibinfo {author} {\bibfnamefont {G.~V.}\ \bibnamefont
  {Shlyapnikov}},\ }\href {http://link.aps.org/doi/10.1103/PhysRevA.67.031601}
  {\bibfield  {journal} {\bibinfo  {journal} {Phys. Rev. A}\ }\textbf {\bibinfo
  {volume} {67}},\ \bibinfo {pages} {031601} (\bibinfo {year}
  {2003})}\BibitemShut {NoStop}%
\bibitem [{\citenamefont {Bertaina}\ and\ \citenamefont
  {Giorgini}(2011)}]{DimerDimer_Scattering_Giorgini}%
  \BibitemOpen
  \bibfield  {author} {\bibinfo {author} {\bibfnamefont {G.}~\bibnamefont
  {Bertaina}}\ and\ \bibinfo {author} {\bibfnamefont {S.}~\bibnamefont
  {Giorgini}},\ }\href
  {https://link.aps.org/doi/10.1103/PhysRevLett.106.110403} {\bibfield
  {journal} {\bibinfo  {journal} {Phys. Rev. Lett.}\ }\textbf {\bibinfo
  {volume} {106}},\ \bibinfo {pages} {110403} (\bibinfo {year}
  {2011})}\BibitemShut {NoStop}%
\bibitem [{\citenamefont {Pricoupenko}\ and\ \citenamefont
  {Pedri}(2010)}]{pricoupenko2010}%
  \BibitemOpen
  \bibfield  {author} {\bibinfo {author} {\bibfnamefont {L.}~\bibnamefont
  {Pricoupenko}}\ and\ \bibinfo {author} {\bibfnamefont {P.}~\bibnamefont
  {Pedri}},\ }\href
  {https://journals.aps.org/pra/abstract/10.1103/PhysRevA.82.033625} {\bibfield
   {journal} {\bibinfo  {journal} {Phys. Rev. A}\ }\textbf {\bibinfo {volume}
  {82}},\ \bibinfo {pages} {033625} (\bibinfo {year} {2010})}\BibitemShut
  {NoStop}%
\bibitem [{\citenamefont {Ngampruetikorn}\ \emph {et~al.}(2013)\citenamefont
  {Ngampruetikorn}, \citenamefont {Parish},\ and\ \citenamefont
  {Levinsen}}]{ngampruetikorn2013}%
  \BibitemOpen
  \bibfield  {author} {\bibinfo {author} {\bibfnamefont {V.}~\bibnamefont
  {Ngampruetikorn}}, \bibinfo {author} {\bibfnamefont {M.~M.}\ \bibnamefont
  {Parish}}, \ and\ \bibinfo {author} {\bibfnamefont {J.}~\bibnamefont
  {Levinsen}},\ }\href {http://stacks.iop.org/0295-5075/102/i=1/a=13001}
  {\bibfield  {journal} {\bibinfo  {journal} {EPL (Europhysics Letters)}\
  }\textbf {\bibinfo {volume} {102}},\ \bibinfo {pages} {13001} (\bibinfo
  {year} {2013})}\BibitemShut {NoStop}%
\bibitem [{\citenamefont {Naidon}\ \emph {et~al.}(2016)\citenamefont {Naidon},
  \citenamefont {Endo},\ and\ \citenamefont
  {Garc{\'\i}a-Garc{\'\i}a}}]{ResonatingGroupMethod_Shimpei}%
  \BibitemOpen
  \bibfield  {author} {\bibinfo {author} {\bibfnamefont {P.}~\bibnamefont
  {Naidon}}, \bibinfo {author} {\bibfnamefont {S.}~\bibnamefont {Endo}}, \ and\
  \bibinfo {author} {\bibfnamefont {A.~M.}\ \bibnamefont
  {Garc{\'\i}a-Garc{\'\i}a}},\ }\href
  {http://stacks.iop.org/0953-4075/49/i=3/a=034002} {\bibfield  {journal}
  {\bibinfo  {journal} {Journal of Physics B: Atomic, Molecular and Optical
  Physics}\ }\textbf {\bibinfo {volume} {49}},\ \bibinfo {pages} {034002}
  (\bibinfo {year} {2016})}\BibitemShut {NoStop}%
\bibitem [{Note1()}]{Note1}%
  \BibitemOpen
  \bibinfo {note} {Strictly speaking, a repulsive Fermi gas will always form a
  superfluid at sufficiently low temperature, due to the effective interactions
  induced by the medium. However, since the critical temperature for
  superfluidity is exponentially suppressed compared to all other energy scales
  in the system, we can consider the ground state to be a Fermi
  liquid}\BibitemShut {NoStop}%
\bibitem [{\citenamefont {Parish}\ \emph {et~al.}(2007)\citenamefont {Parish},
  \citenamefont {Marchetti}, \citenamefont {Lamacraft},\ and\ \citenamefont
  {Simons}}]{parish2007}%
  \BibitemOpen
  \bibfield  {author} {\bibinfo {author} {\bibfnamefont {M.~M.}\ \bibnamefont
  {Parish}}, \bibinfo {author} {\bibfnamefont {F.~M.}\ \bibnamefont
  {Marchetti}}, \bibinfo {author} {\bibfnamefont {A.}~\bibnamefont
  {Lamacraft}}, \ and\ \bibinfo {author} {\bibfnamefont {B.~D.}\ \bibnamefont
  {Simons}},\ }\href {http://dx.doi.org/10.1038/nphys520} {\bibfield  {journal}
  {\bibinfo  {journal} {Nature Phys.}\ }\textbf {\bibinfo {volume} {3}},\
  \bibinfo {pages} {124} (\bibinfo {year} {2007})}\BibitemShut {NoStop}%
\bibitem [{\citenamefont {He}\ \emph {et~al.}(2015)\citenamefont {He},
  \citenamefont {L\"u}, \citenamefont {Cao}, \citenamefont {Hu},\ and\
  \citenamefont {Liu}}]{he2015quantum}%
  \BibitemOpen
  \bibfield  {author} {\bibinfo {author} {\bibfnamefont {L.}~\bibnamefont
  {He}}, \bibinfo {author} {\bibfnamefont {H.}~\bibnamefont {L\"u}}, \bibinfo
  {author} {\bibfnamefont {G.}~\bibnamefont {Cao}}, \bibinfo {author}
  {\bibfnamefont {H.}~\bibnamefont {Hu}}, \ and\ \bibinfo {author}
  {\bibfnamefont {X.-J.}\ \bibnamefont {Liu}},\ }\href
  {http://link.aps.org/doi/10.1103/PhysRevA.92.023620} {\bibfield  {journal}
  {\bibinfo  {journal} {Phys. Rev. A}\ }\textbf {\bibinfo {volume} {92}},\
  \bibinfo {pages} {023620} (\bibinfo {year} {2015})}\BibitemShut {NoStop}%
\bibitem [{\citenamefont {Altland}\ and\ \citenamefont
  {Simons}(2010)}]{CondensedMatter_FieldTheory_Altland}%
  \BibitemOpen
  \bibfield  {author} {\bibinfo {author} {\bibfnamefont {A.}~\bibnamefont
  {Altland}}\ and\ \bibinfo {author} {\bibfnamefont {B.~D.}\ \bibnamefont
  {Simons}},\ }\href@noop {} {\emph {\bibinfo {title} {Condensed matter field
  theory}}}\ (\bibinfo  {publisher} {Cambridge University Press},\ \bibinfo
  {year} {2010})\BibitemShut {NoStop}%
\bibitem [{\citenamefont {Cooper}(1956)}]{CooperPair_Original_Cooper}%
  \BibitemOpen
  \bibfield  {author} {\bibinfo {author} {\bibfnamefont {L.~N.}\ \bibnamefont
  {Cooper}},\ }\href {http://link.aps.org/doi/10.1103/PhysRev.104.1189}
  {\bibfield  {journal} {\bibinfo  {journal} {Phys. Rev.}\ }\textbf {\bibinfo
  {volume} {104}},\ \bibinfo {pages} {1189} (\bibinfo {year}
  {1956})}\BibitemShut {NoStop}%
\bibitem [{\citenamefont {Gorshkov}\ \emph {et~al.}(2010)\citenamefont
  {Gorshkov}, \citenamefont {Hermele}, \citenamefont {Gurarie}, \citenamefont
  {Xu}, \citenamefont {Julienne}, \citenamefont {Ye}, \citenamefont {Zoller},
  \citenamefont {Demler}, \citenamefont {Lukin},\ and\ \citenamefont
  {Rey}}]{gorshkov2010two}%
  \BibitemOpen
  \bibfield  {author} {\bibinfo {author} {\bibfnamefont {A.}~\bibnamefont
  {Gorshkov}}, \bibinfo {author} {\bibfnamefont {M.}~\bibnamefont {Hermele}},
  \bibinfo {author} {\bibfnamefont {V.}~\bibnamefont {Gurarie}}, \bibinfo
  {author} {\bibfnamefont {C.}~\bibnamefont {Xu}}, \bibinfo {author}
  {\bibfnamefont {P.}~\bibnamefont {Julienne}}, \bibinfo {author}
  {\bibfnamefont {J.}~\bibnamefont {Ye}}, \bibinfo {author} {\bibfnamefont
  {P.}~\bibnamefont {Zoller}}, \bibinfo {author} {\bibfnamefont
  {E.}~\bibnamefont {Demler}}, \bibinfo {author} {\bibfnamefont
  {M.}~\bibnamefont {Lukin}}, \ and\ \bibinfo {author} {\bibfnamefont
  {A.}~\bibnamefont {Rey}},\ }\href
  {http://www.nature.com/nphys/journal/v6/n4/full/nphys1535.html} {\bibfield
  {journal} {\bibinfo  {journal} {Nature Phys.}\ }\textbf {\bibinfo {volume}
  {6}},\ \bibinfo {pages} {289} (\bibinfo {year} {2010})}\BibitemShut {NoStop}%
\bibitem [{\citenamefont {Zhang}\ \emph {et~al.}(2015)\citenamefont {Zhang},
  \citenamefont {Cheng}, \citenamefont {Zhai},\ and\ \citenamefont
  {Zhang}}]{PhysRevLett.115.135301}%
  \BibitemOpen
  \bibfield  {author} {\bibinfo {author} {\bibfnamefont {R.}~\bibnamefont
  {Zhang}}, \bibinfo {author} {\bibfnamefont {Y.}~\bibnamefont {Cheng}},
  \bibinfo {author} {\bibfnamefont {H.}~\bibnamefont {Zhai}}, \ and\ \bibinfo
  {author} {\bibfnamefont {P.}~\bibnamefont {Zhang}},\ }\href
  {https://link.aps.org/doi/10.1103/PhysRevLett.115.135301} {\bibfield
  {journal} {\bibinfo  {journal} {Phys. Rev. Lett.}\ }\textbf {\bibinfo
  {volume} {115}},\ \bibinfo {pages} {135301} (\bibinfo {year}
  {2015})}\BibitemShut {NoStop}%
\bibitem [{\citenamefont {H\"ofer}\ \emph {et~al.}(2015)\citenamefont
  {H\"ofer}, \citenamefont {Riegger}, \citenamefont {Scazza}, \citenamefont
  {Hofrichter}, \citenamefont {Fernandes}, \citenamefont {Parish},
  \citenamefont {Levinsen}, \citenamefont {Bloch},\ and\ \citenamefont
  {F\"olling}}]{PhysRevLett.115.265302}%
  \BibitemOpen
  \bibfield  {author} {\bibinfo {author} {\bibfnamefont {M.}~\bibnamefont
  {H\"ofer}}, \bibinfo {author} {\bibfnamefont {L.}~\bibnamefont {Riegger}},
  \bibinfo {author} {\bibfnamefont {F.}~\bibnamefont {Scazza}}, \bibinfo
  {author} {\bibfnamefont {C.}~\bibnamefont {Hofrichter}}, \bibinfo {author}
  {\bibfnamefont {D.~R.}\ \bibnamefont {Fernandes}}, \bibinfo {author}
  {\bibfnamefont {M.~M.}\ \bibnamefont {Parish}}, \bibinfo {author}
  {\bibfnamefont {J.}~\bibnamefont {Levinsen}}, \bibinfo {author}
  {\bibfnamefont {I.}~\bibnamefont {Bloch}}, \ and\ \bibinfo {author}
  {\bibfnamefont {S.}~\bibnamefont {F\"olling}},\ }\href
  {https://link.aps.org/doi/10.1103/PhysRevLett.115.265302} {\bibfield
  {journal} {\bibinfo  {journal} {Phys. Rev. Lett.}\ }\textbf {\bibinfo
  {volume} {115}},\ \bibinfo {pages} {265302} (\bibinfo {year}
  {2015})}\BibitemShut {NoStop}%
\bibitem [{\citenamefont {Pagano}\ \emph {et~al.}(2015)\citenamefont {Pagano},
  \citenamefont {Mancini}, \citenamefont {Cappellini}, \citenamefont {Livi},
  \citenamefont {Sias}, \citenamefont {Catani}, \citenamefont {Inguscio},\ and\
  \citenamefont {Fallani}}]{pagano2015}%
  \BibitemOpen
  \bibfield  {author} {\bibinfo {author} {\bibfnamefont {G.}~\bibnamefont
  {Pagano}}, \bibinfo {author} {\bibfnamefont {M.}~\bibnamefont {Mancini}},
  \bibinfo {author} {\bibfnamefont {G.}~\bibnamefont {Cappellini}}, \bibinfo
  {author} {\bibfnamefont {L.}~\bibnamefont {Livi}}, \bibinfo {author}
  {\bibfnamefont {C.}~\bibnamefont {Sias}}, \bibinfo {author} {\bibfnamefont
  {J.}~\bibnamefont {Catani}}, \bibinfo {author} {\bibfnamefont
  {M.}~\bibnamefont {Inguscio}}, \ and\ \bibinfo {author} {\bibfnamefont
  {L.}~\bibnamefont {Fallani}},\ }\href
  {https://journals.aps.org/prl/abstract/10.1103/PhysRevLett.115.265301}
  {\bibfield  {journal} {\bibinfo  {journal} {Phys. Rev. Lett.}\ }\textbf
  {\bibinfo {volume} {115}},\ \bibinfo {pages} {265301} (\bibinfo {year}
  {2015})}\BibitemShut {NoStop}%
\end{thebibliography}%

\end{document}